\documentclass[manuscript,screen]{acmart}

\AtBeginDocument{%
  \providecommand\BibTeX{{%
    \normalfont B\kern-0.5em{\scshape i\kern-0.25em b}\kern-0.8em\TeX}}}

\setcopyright{acmcopyright}
\copyrightyear{2018}
\acmYear{2018}
\acmDOI{XXXXXXX.XXXXXXX}

\usepackage{times} 
\usepackage{blindtext, graphicx}
\usepackage{url}
\usepackage{multirow} 
\usepackage{tabu}
\usepackage{colortbl}
\usepackage{wrapfig}
\usepackage{tikz}
\usepackage{wrapfig}
\usepackage{adjustbox}
\usepackage{fixltx2e}

\usepackage{pifont}
\usepackage{tcolorbox}
\usepackage{svg}

\pdfoutput=1
\usepackage{tikz}

\usepackage{graphicx}

\usepackage{microtype}

\newcommand{\bi}{\begin{itemize}}

\newcommand{\ei}{\end{itemize}}

\usepackage[tikz]{bclogo}
{\noindent\begin{minipage}[c]{\linewidth}%
\begin{bclogo}[couleur=gray!25,%
arrondi=0.1,%
logo=\bctrombone,%
ombre=true]{{\small ~#1}}}%
{\end{bclogo}\end{minipage}\vspace{2mm}}

\newcommand{\tbl}[1]{Table~\ref{tbl:#1}}
\newcommand{\fig}[1]{Figure~\ref{fig:#1}}
\usepackage{balance}
\usepackage{pifont}
\definecolor{Gray}{rgb}{0.88,1,1}
\definecolor{Gray}{gray}{0.85}
\definecolor{lightgray}{gray}{0.8}

\usepackage{caption}
\usepackage{subcaption}

\usepackage[linesnumbered,ruled,vlined]{algorithm2e}
\SetKwProg{Fn}{Function}{}{}

\usepackage{amsmath}

\usepackage{framed}
\usepackage{tikz}
\usetikzlibrary{shadows}
\theoremstyle{break}

\tikzstyle{thmbox} = [rectangle, rounded corners, draw=black,
fill=Gray!20, drop shadow={fill=black, opacity=1}]

\newcommand{\tion}[1]{\S\ref{tion:#1}}

\newcommand{\respto}[1]{}

\newcommand{\BLUE}{\color{black}}
\newcommand{\BLACK}{\color{black}}
\newcommand{\ORANGE}{\color{orange}}

\usepackage{color}
\usepackage{xspace}
\definecolor{ScarletRed}{rgb}{0.80,0.00,0.00}

\newcommand{\hil}{\cellcolor[gray]{.9}}

\newcommand{\all}{\cellcolor[HTML]{85C1E9}}

\newcommand{\early}{\cellcolor[HTML]{A3E4D7}}
\newcommand{\tca}{\cellcolor[HTML]{F8F9F9}}
\newcommand{\bellwether}{\cellcolor[HTML]{F8F9F9}}

\definecolor{recency}{HTML}{F4D03F}
\definecolor{all}{HTML}{85C1E9}
\definecolor{early}{HTML}{A3E4D7}
\definecolor{white}{HTML}{FFFFFF}
\definecolor{tca}{HTML}{F8F9F9}
\definecolor{bellwether}{HTML}{F8F9F9}
\definecolor{ebell}{HTML}{A3E4D7} 

\definecolor{etwo}{HTML}{B0E0E6}
\definecolor{ebtwo}{HTML}{00B1BC}

\definecolor{lower}{HTML}{FF9933}
\definecolor{higher}{HTML}{82E0AA}

\begin{document}

\title{Assessing the  Early Bird Heuristic (for Predicting Project Quality)}

\author{Shrikanth N.C.}
\email{nc.shrikanth@gmail.com}
\orcid{0000-0002-8983-7733}
\author{Tim Menzies}
\email{timm@ieee.org}
\affiliation{%
  \institution{North Carolina State University}
  \streetaddress{Engineering Building II - Campus Box 8206, 890 Oval Dr}
  \city{Raleigh}
  \state{North Carolina}
  \country{USA}
  \postcode{27606}
}




\begin{abstract}
Before researchers rush to reason across all available data or try complex methods,
perhaps it is prudent to first check for simpler alternatives. Specifically, if the historical data 
has the most information in some small region, perhaps a model learned from that region
would suffice for the rest of the project.

To support this claim, we offer a case study with 240   projects, where we find
that the information in those projects ``clumpe'' towards the earliest parts of the project.
A quality prediction model learned from just the first 150 commits works as well, or better
than state-of-the-art alternatives. Using just this "early bird"  data, we can build models
very quickly and very early in the  project life cycle. Moreover, using this early bird method, we have shown that a simple model (with just a few features) generalizes to hundreds of  projects.

Based on this experience, we doubt that prior work on generalizing quality
 models may have needlessly complicated an inherently simple process.
Further, prior work that focused on later-life cycle data needs to be revisited since their 
conclusions were drawn from relatively uninformative regions. 

Replication note: all our data and scripts are available here: https://github.com/snaraya7/early-bird
\end{abstract}

\ccsdesc[500]{Software and its engineering~Maintaining software}

\keywords{quality prediction,  defects,  early, data-lite}

 \maketitle

\section{Introduction}\label{tion:introduction}

\begin{figure*}[!b]
\begin{center}

\includegraphics[width=\linewidth]{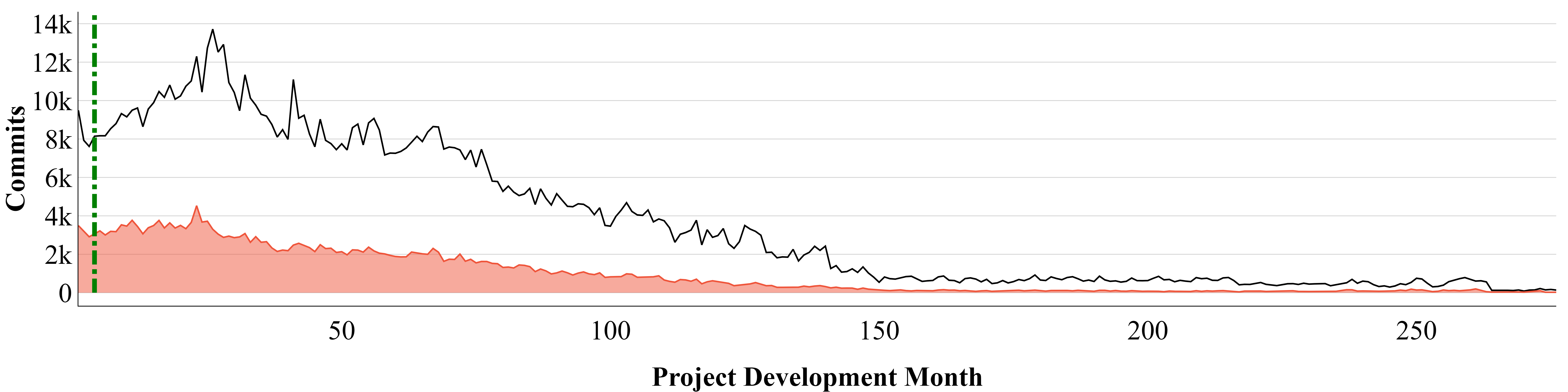}

{\bf Figure~\ref{fig:whale}.A}: 155  popular GitHub projects (\#stars$>1000$). 
Data from 1.2 million commits.

\includegraphics[width=1\linewidth]{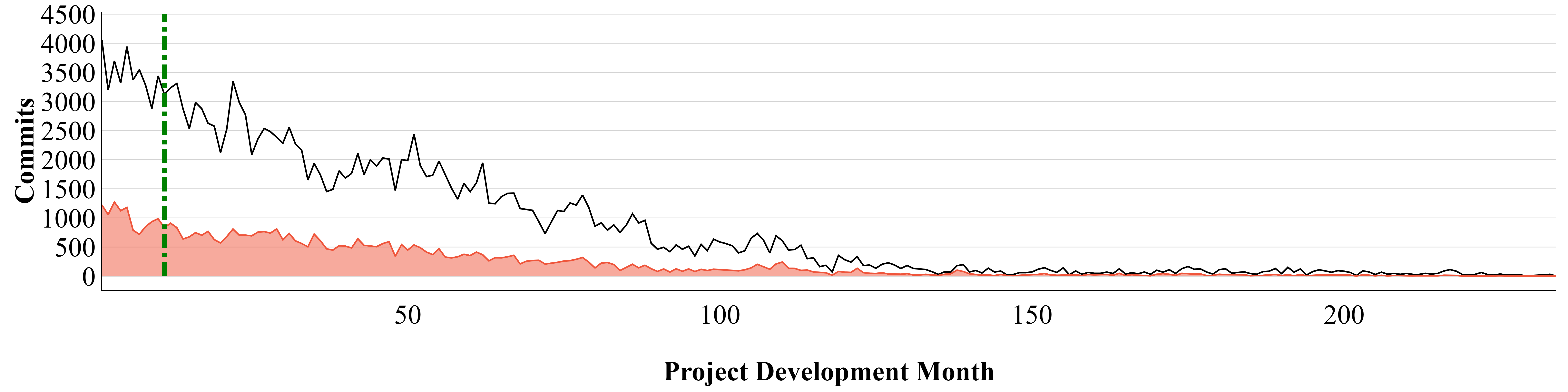}

{\bf Figure~\ref{fig:whale}.B}:  85 unpopular GitHub projects (\#stars$<1000$).
Data from 
253,289 commits.

\end{center}

\caption{
Most defective commits occur
early in the life cycle.
Black:Red  = Clean:Defective commits. In this paper, we compare (a) models learned up to the vertical green (dotted) line to (b)   models learned using more data.}\label{fig:whale}
\end{figure*}

In software defect prediction, researchers often fail to try simpler methods before trying complex analytics~\cite{icse21,zhou2018far}. For example:  
\bi

\item ``For many new projects we may not have enough historical data to train
prediction models''~\cite{rahman2012recalling}

\item ``At least for defect prediction, it is no longer enough to just run a data miner and present the result without conducting a tuning optimization study.''~\cite{fu2016tuning}

\item ``Many research studies have shown that ensemble learning can
achieve much better classification performance than a single classifier''~\cite{yang2017tlel}

\ei

While we do not doubt those results in that context, we show in this work that \BLUE \respto{3M1} we can simplify software defect prediction if \BLACK we focus more on the quality of the training data than on applying various software analytic techniques. Furthermore, this paper shows that many of the prevalent methods used to build defect predictors are needless and can be `simplified' to a large extent with knowledge-rich early project data. 

By simplified we mean:
\bi
\item Stable conclusions (no need to update defect predictors for every new project release). 
\item Less run-time (With fewer project data we can find and build relevant predictors faster). 
\item Better explanations (With fewer features we can communicate the oracles (predictors) decisions to stakeholders effectively). 
\ei

  



\respto{2C}\BLUE In our lab exercise, we have sometimes seen unsatisfactory results from such a complex approach. For example, once we tried learning defect attributes 
 from 700,000+ commits.  The  web slurping required for that process took nearly 500 CPU days, using five machines with 16 cores, over seven days (since the data from those projects had to be cleaned, massaged, and transformed into some standard format). Within that data space, we found significant differences in the models learned from different parts of the data. So even after all that work and over a year of CPU, we were unable to report a stable conclusion to our business users. \BLACK

This kind of experience motivate
us to ask the question ``when is enough data just enough to build effective defect predictors?''
Our answer to this question comes from 
a previously unreported effect is shown in  
\fig{whale}.
As shown in this figure,
when we look at the  
percent of buggy commits in GitHub projects, a  remarkable pattern
emerges. Specifically, 
most of the buggy commits
occur earlier in the life cycle.

This observation prompted an investigation of the ``early bird'' heuristic (or the ``early-data-lite'' sampling method) that just uses early life cycle data to predict defects.
The literature review of \S\ref{secrelated} shows that, surprisingly, this
 approach to software defect prediction has been overlooked by prior work~\cite{icse21}. 
This is a   significant oversight since,
 in 240 GitHub projects, we can show that 
defect predictors learned from the first 150 commits 
work as well, or better, than state-of-the-art 
alternatives.
 
Our prior work showed early-data-lite ($E$) within project supervised defect prediction models performed  statistically better or on par with those that were built using either recent ($RR$) or abundant commits ($ALL$)~\cite{icse21}. In this work, we exploit the early-data-lite ($E$) method to answer \BLUE \respto{1M1} four \BLACK research questions (RQs), they are:
\bi
\item[]  RQ1: Can we build early software defect prediction models from unpopular projects?
\item[] RQ2: Can we build early software defect prediction models with fewer features?
\item[] RQ3: \BLUE \respto{3M2}Can we build \BLACK early software defect prediction models by transferring early life cycle data from other projects?
\item [] RQ4: Do complex methods supersede early software defect prediction models?
\ei

Results of \BLUE \respto{1M1} four \BLACK RQs in \tion{results} confirm that the software defect prediction model built from \respto{2H} \BLUE another project (cross)  from the same organization using the early bird heuristic can identify defects in local projects having fewer \BLACK commits. Notably, the early bird (discussed later in \tion{early-heuristic}) heuristic offers the following benefits: a) only the first 150 commits are considered for training; therefore the cross-project can be young (only a few months old), b) we find using only two size-based features to build defect prediction models was sufficient, c) the cross-project need not be well maintained (unpopular) and d) the results also indicate that much of the augmenting machine learning practices such as hyper-parameter optimization and ensemble learning were needless.

Overall, the  contributions of this paper are to show:
\begin{itemize}
\item The information (like software defects) within projects   may not be evenly distributed across the  life cycle.
For such data, it can be very useful to adopt a ``early-data-lite'' approach.
\item  
For example,    early life cycle data finds  simple models  (with only two features) that generalize across
hundreds of  projects. 
Such models
 can be built much faster than traditional methods (weeks versus months of CPU time).
\item
So before researchers use all available data,
they need to first check that their buggy commit data occur at equal frequency across the life cycle.
We say this since
much prior work
 on methods for learning from multiple projects~\cite{menzies2011local,
li2012sample,
ma2012transfer,
he2012investigation,
menzies2012local,
bettenburg2012think,
rahman2012recalling,
turhan2013empirical,
nam2013transfer,
peters2013better,
canfora2013multi,
peters2013balancing,
herbold2013training,
he2013learning,
rahman2013and,
fukushima2014empirical,
panichella2014cross,
zhang2014towards,
ryu2015hybrid,
chen2015negative,
zhang2015empirical,
peters2015lace2,
canfora2015defect,
jing2015heterogeneous,
he2015empirical,
zhang2016cross,
kamei2016studying,
yang2016effort,
ryu2016value,
xia2016hydra,
jing2016improved,
ryu2016effective,
zhang2016towards,
krishna2016too,
wang2016automatically,
ryu2017transfer,
nam2017heterogeneous,
ni2017cluster,
zhou2018far,
chen2018multi,
hosseini2018benchmark,
krishna2018bellwethers,
li2018cost,
wang2018deep,
wu2018cross,
dam2018deep,
huang2019revisiting,
chen2019multiview,
chen2019software,
liu2019two}   needlessly complicated an inherently simple
process. 
\end{itemize}








\begin{table}[!t]
\caption{Table of Important Acronyms}
\label{tbl:acronyms}
\centering
{\scriptsize

\begin{tabular}{l|l}
\rowcolor[HTML]{EFEFEF} 
\textbf{Acronym} & \textbf{Abbreviation} \\
\textbf{AUC} & Area under the receiver operating characteristic curve \\
\textbf{CFS} & Correlation-based Feature Selection \\
\textbf{DODGE} & Optimizer proposed by Agrawal et al. in~\cite{agrawal2019dodge} \\
\textbf{DT} & Decision Tree \\
\textbf{HPO} & Hyperparameter optimization \\
\textbf{HYPEROPT} & Optimizer proposed by Bergstra et al.~\cite{Bergstra:2011} \\
\textbf{IFA} & Initial Number of False Alarms \\
\textbf{KNN} & k-nearest neighbors algorithm \\
\textbf{LA and LT} & Refer to features list in \tbl{features} \\
\textbf{LR} & Logistic Regression \\
\textbf{MCC} & Matthews correlation coefficient \\
\textbf{NB} & Naive Bayes classifier \\
\textbf{PF} & False Alarm Rate \\
\textbf{RF} & Random forest \\
\textbf{SMOTE} & Synthetic Minority Over-sampling Technique~\cite{chawla2002smote} \\
\textbf{SVM} & Support vector machines \\
\textbf{SZZ} & Sliwerski Zimmerman Zellar Algorithm~\cite{cite_szz_original}\\
\textbf{TCA} & Transfer Component Analysis~\cite{nam2013transfer} \\
\textbf{TLEL} & Two-layer Ensemble Learning Algorithm~\cite{yang2017tlel} \\
\textbf{TPE} & Tree-structured Parzen Estimator~\cite{Bergstra:2011}
\end{tabular}
}
\end{table}

The rest of this paper  explains our method;
and offers experimental evidence that this   early-bird method
works better than sophisticated methods like classifier tuning, ensemble and notably made  transfer learning algorithms work  simpler, faster, and more stable. 
This paper uses the abbreviations
of \tbl{acronyms}.

Before all that, we digress to address the obvious objections to our conclusion.
The results presented here are only for software defect
prediction. In future work,
we need to test if our
results hold for other domains.

\section{Connection to Prior Work}
\subsection{Initial Report}
Previously, we have reported
the \fig{whale}. A results
at ICSE'21~\cite{icse21}\footnote{For reviewers, we note that that paper is available on-line
at https://arxiv.org/pdf/2011.13071.pdf}.
We calculate that only \fig{whale}.A  and  5 pages of this text (e.g.~\S\ref{manage}) come from that prior work. 

As to the semantic difference between this paper to prior work,
that prior study was  more limited   since it did not have 
the additional data of 
\fig{whale}.B. 
Also,
that prior study
only ran some within-project sampling methods on the \fig{whale}.A data. This paper
takes the additional step
of comparing our methods
to transfer learning, classifier tuning, and ensemble methods.
Further, previously, we did not report the model learned via this method. We show here that a simple  model (that uses just a few variables)
can be learned from a few early life cycle samples. 
Lastly,  as shown by the new results of this paper in \tbl{rq1}, \tbl{rq2}, \tbl{rq3}, and \tbl{rq4} we can out-perform
that results from that prior work. Hence, it is important to ingest the results of our prior work first, especially to understand why we omit certain treatments in our experiments in this work.

\subsection{Early Bird Heuristic}\label{tion:early-heuristic}
In our prior work, we find that much of the software engineering (SE) project activity occurs during the early periods of the software project  life-cycle~\cite{icse21}.  We used that observation as a heuristic to answer if defect prediction models can stop learning early. Interestingly, we found that defect prediction models that learned from just 150 commits (which occurred in less than four months of the project) yield \respto{1A}\BLUE statistically the same predictive performance as models  trained with recent or all available project data (commits) \BLACK.  In this work, we label this effect as the `early-bird' heuristic (or $E$).

Software defect predictors built in the literature or practice using all available data are data-hungry or those that use recent project data may be labeled as late data. The defect prediction models built using the `early-bird' heuristic are early-data-lite as they use only a  tiny portion (4\%) of the early project data~\cite{icse21}.  Data-hungry models are frequently re-trained and this gives rise to the conclusion-instability issue~\cite{krishna2018bellwethers}. In other words, retraining would cause the models to behave differently for every project release (e.g, A commit classified as defective in release `X' may be classified as clean in release `X+1'). Early-data-lite models are `never retrained' as they always use the fixed 150 early project commits to building defect predictors.  \BLUE

\respto{1B} \section{Motivation}\label{tion:background} 
\BLACK
\subsection{Defect Prediction}

The case studies of this paper are based on software defect prediction. Hence, before
doing anything else, we need to introduce that research area.

Fixing software defects is not cheap~\cite{hailpern2002software}. 
Accordingly, for decades, SE researchers have been devising numerous ways to predict software quality before deployment. One of the oldest studies was made in 1971 by Akiyama using size-based defect predictions for a system developed at Fujitsu, Japan~\cite{akiyama1971example}. This approach remains popular today. A 2018 survey of 395 practitioners
from 33 countries and five continents [20] found that over
90\% of the respondents were willing to adopt software defect prediction techniques~\cite{wan2018perceptions}.

 Software defect prediction
uses  data miners to input static code attributes and 
output models that  predict where
  the   code  probably contains   most bugs~\cite{ostrand2005predicting,menzies2006data}.
  The models learned in this way are very effective and relatively
  cheap to build:
  \begin{itemize}
   
\item {\em Effective}:
Misirili et al. ~\cite{misirli2011ai}, and Kim et al.~\cite{kim2011empirical} report
considerable cost savings when such predictors are used in guiding industrial quality assurance processes. 
\item {\em Relatively simpler to implement}
Also, Rahman et al.~\cite{10.1145/2568225.2568269} show that such predictors are competitive with more elaborate approaches. For example, they note
that static code analysis tools can have expensive licenses that need
to be updated after any major language upgrade. Defect predictors, on the other hand, can be quickly implemented via some lightweight parsing of commit-level metrics that are programming language agnostic.
\end{itemize}

Our prior work~\cite{icse21} proposed numerous extensions that improve the validity and scope of the ``early-data-lite'' method. Therefore, in this work we continue to assess the efficacy and applicability of the ``early-data-lite'' method broadly on two active areas in software defect prediction, specifically:

\bi
\item \textbf{Data:} Transfer Learning (Cross-project defect prediction)
\item \textbf{Technique:} Tuning and Ensemble methods 
\ei

\subsection{Data: Transfer Learning}\label{tion:data_sampling}


Defect predictors are learned from project data. What happens if there is not
enough data to learn those models? This is an especially acute problem for newer projects
(and in the absence of historical data in some legacy projects~\cite{briand2002assessing}). 

In such scenarios, practitioners and researchers might identify matured projects that share some similarities to their local projects. Once found, then lessons learned could
be {\em transferred} from the older to the new project. There are kinds of transfer:
\begin{itemize}
\item
\textit{Cross}: {\em cross-project defect prediction}. Lessons   learned from {\em other} projects
and applied to {\em this} project.
\item
\textit{Within}: {\em within-project defect prediction}. Using data from {\em this} project, lessons learned from prior experience  are used to make predictions about later life cycle development.

\end{itemize}
To say the least, transfer learning is a very active research area in SE. We can find more than 1,000 articles in the last five years alone 
(found using the query  "cross-project defect prediction,"~\footnote{Queried https://scholar.google.co.in/ in 2020}). By our count,
within that corpus, there have been at least two dozen transfer learning methods ~\cite{amasaki2020cross}.
Interesting methods
evolved in that research include:
\bi
\item Heterogeneous transfer that lets data expressed
indifferent formats transferred from project to projects~\cite{nam2017heterogeneous};

\item Temporal transfer learning, which is a within-project
defect prediction (\textit{Within}) tool where earlier life cycle data is used
to make predictions later in the life cycle~\cite{kocaguneli2012}.
\ei

Our reading of the literature is that, apart from our own research,
the prior state-of-the-art in the SE literature is Nam et al.'s $TCA+$ (Transfer Component Analysis) method~\cite{nam2013transfer}. For a list of important abbreviations used in this paper see \tbl{acronyms}.
Given data from a source and a target  project, TCA strives to ``align''
the source and target data via dimensionality rotation, expansion, and
contraction. TCA+ is an extension to basic TCA that used automatic
methods to find normalization options for TCA.

\subsection{Technique : Tuning and Ensemble}\label{tuning_ensemble}
Classification algorithms can be trained to classify a project commit as defective. Studies have shown  their predictive performance can depend upon the set of hyper-parameter they are initially configured~\cite{fu2016tuning,agrawal2019dodge}.

For example, the   k-nearest neighbors ($KNN$) that we explore in this study are available in the widely used scikit-learn~\cite{scikit-learn} machine learning library. And $KNN$ is set to the following default parameters based on standard machine learning literature as follows:
\begin{center}
$n\_{\mathit neighbors}=5, *, {\mathit weights}={\mathit uniform}, {\mathit algorithm}={\mathit auto}, {\mathit leaf}\_{\mathit size}=30$,\\ $p=2,  {\mathit metric}={\mathit minkowski}, {\mathit metric}\_{\mathit params}={\mathit None}, n\_{\mathit jobs}={\mathit None}$
\end{center}

However, as seen from the numerous parameters available for $KNN$, there is a huge parameter space of options the $KNN$ can be tried and tested (tuned). Therefore most machine learning algorithms' hyper-parameters can be tuned using training data before being tested on project releases in our case.

There are numerous ways to find the right set of parameters for a classifier given the training data. But the decision predominantly comes with the run-time cost. For example, searching through all available options `Manual Search' may not terminate. One may try `Random Search' with a terminating condition but the probability of finding the near-best parameter may be low. Fu et al. \cite{fu2016tuning} explored `grid-search' a baseline hyper-parameter approach in the field of machine learning~\cite{fu2016tuning} but found it to be very slow for software defect prediction. Recently (2019) Agrawal et al. proposed a novel optimizer that terminates quickly after finding near-optimal hyper-parameters for defect predictors~\cite{agrawal2019dodge}. More details on DODGE will be presented in \tion{dodge}. Therefore this paper will explore DODGE with early methods.

Another avenue of complex approach is the use of the ensemble method, where the philosophy is why use just `one' when `more' is better.  \BLUE  \respto{1M2} Numerous studies have shown defect predictors can be improved significantly \respto{3M3} when an ensemble  \BLACK of classifiers were used rather than just one~\cite{yang2017tlel,wang2013using,he2009learning}. In 2017, Yang et al. proposed a two-layer ensemble approach for software defect prediction called TLEL (Two-layer Ensemble Learner) that showed promising improvements when compared to  predictors with a single classifier. \tion{tlel} elucidates its operation and this paper will explore TLEL as part of the complex methods to test the efficacy of early methods.

\subsection{Issues with current approaches}

Nevertheless, just because a technique is popular does not mean that it should be recommended. Our reading of the literature is most  recent SE analytics papers have taken a complex  approach (e.g. see the quotes in our prior work~\cite{icse21}) and use state-of-the-art analytics without testing with simpler alternatives~\cite{zhou2018far}. We argue there that  it can be useful to try early-data-lite before adopting complex  techniques that demand more data, delay analytics, do not offer explanations, and overutilize system resources. Because, for one thing, all that data might not be available. The availability of more data in an industrial setting is not assured. Further, it may not be useful to learn from more data.  Proprietary data may not be readily available for practitioners to build predictors for their local projects. 
Zimmermann et al. showed that transferring predictors from the same domain does not guarantee quality predictions~\cite{zimmermann2009cross}.
Finally, due to privacy concerns, teams even within the same organization may not readily make their matured project available for others to use~\cite{peters2013balancing}. 

Another thing, several  researchers have
reported that a complex approach can be problematic:
\bi
\item In our introduction, we reported on our own issues seen when learning from 700,000+  GitHub commits.
\item At her ESEM'11 keynote address, Elaine Weyuker questioned whether she will ever have the option to make the AT\&T information public~\cite{weyuker2008too}.
\item In over 30 years of COCOMO effort,  Boehm could share
cost estimation data from only 200 \respto{3E} \BLUE projects~\cite{boehm2005cocomo}.\BLACK
\ei

Next, applying techniques discussed in \tion{sampling_methods} comes with a CPU run-time cost. When  time to time researchers have strived and endorsed to look for simpler alternatives it is important that we explore them.


\section{Related Work}\label{secrelated}

In our prior \respto{3M4} \BLUE work \BLACK, we showed that historical data within the project were used in large quantities to build defect predictors that were needless~\cite{icse21}. And as promised in our prior work in this work  we check the validity of our prior conclusion
beyond within-project software defect prediction and test the efficacy of early methods in various contexts such as:  

\bi
\item Unpopular SE projects
\item Transfer learning scenarios
\item Hyper-parameter optimization and Ensemble approach
\ei

\respto{1C} \BLUE One reason to explore early life cycle data reasoning is that it has not been done before (except our prior work that focuses on within-project data to build predictors). \respto{2I} Fenton et al. (2008) examined the use of human judgment rather than domain data to handcraft a causal model to predict residual defects (defects discovered during operational usage or independent testing), however, this required extensive work (2 years) and we do not explore this path~\cite{fenton2008effectiveness}. Zhang and Wu demonstrated in 2010 that fewer programs sampled from an entire space of programs (covering the entire project life-cycle) can be used to estimate project quality~\cite{zhang2010sampling}. The difference here is that we sample them `early' in the project life-cycle and not the entire project history. Rahman et al. endorse employing a substantial sample size in defect prediction models to reduce bias~\cite{rahman2013sample}. While we do not deny bias in defect prediction data sets, we find that the performance of our proposed early data-lite approach is comparable to that of prior approaches. Arokiam and Jeremy looked into bug severity prediction~\cite{Arokiam2020}. They demonstrate that data transferred from other projects can be used to predict bug severity early in project development. Similar to Arokiam and Jeremy's work, Sousuke examined Cross-Version defect prediction (CVDP) using Cross-Project Defect Prediction (CPDP) data in 2020 in ~\cite{amasaki2020cross}. Their study with 41 releases suggests that the most recent project release was still superior to the majority of CPDP approaches. However, in contrast to Sousuke, we provide contrary evidence in this work. Notably, we evaluate our strategy using more than a thousand releases and seven performance metrics. 

We expanded the scope of early life cycle data reasoning in the subsequent \tion{ls} through a literature survey on other prominent defect prediction approaches specifically transfer learning, hyper-parameter optimization, and ensemble approach. Notably, these methods demand more data and computing resources making it difficult to predict defects early in the project life cycle.  

\BLACK
We find that much of the SE transfer learning methods~\cite{turhan2013empirical,
nam2013transfer,
peters2013better,
canfora2013multi,
peters2013balancing,
fukushima2014empirical,
panichella2014cross,
ryu2015hybrid,
chen2015negative,
zhang2015empirical,
peters2015lace2,
canfora2015defect,
jing2015heterogeneous,
kamei2016studying,
yang2016effort,
ryu2016value,
xia2016hydra,
jing2016improved,
ryu2016effective,
zhang2016towards,
krishna2016too,
ryu2017transfer,
nam2017heterogeneous,
ni2017cluster,
zhou2018far,
chen2018multi,
hosseini2018benchmark,
krishna2018bellwethers,
li2018cost,
wang2018deep,
huang2019revisiting} and Tuning or ensemble methods ~\cite{bowes2018software,
tan2015online,
xia2016hydra,
laradji2015software,
tantithamthavorn2016automated,
wang2018deep,
ryu2016value,
zhou2019improving,
tantithamthavorn2018impact,
rhmann2020software,
kamei2016studying,
yang2017tlel,
chekam2020selecting,
song2018comprehensive,
bird2011don,
ghotra2015revisiting,
wang2013using,
ryu2015hybrid,
panichella2014cross,
jing2016improved,
ryu2017transfer,
peters2013better,
wang2016multiple,
sun2012using,
siers2015software,
pandey2020bpdet,
huda2018ensemble,
xu2019software,
tong2018software,
pascarella2019fine,
rathore2017towards,
kamei2016studying,
lam2017bug,
ye2014learning,
scandariato2014predicting,
agrawal2018better,
agrawal2018wrong,
zhang2016towards,
agrawal2019dodge,
jiang2013personalized,
fu2016tuning,
li2017software,
wang2016automatically,
liu2019two} could be characterized as ``late-data'' (or data-hungry)
\BLUE since they transferred all available data from one or more projects to build their predictors supporting our claim that this space has not been explored before. \BLACK
 
As to other \BLUE \respto{1M3}\respto{3M5}comparison studies, 
we \BLACK explore transfer learning
and hyper-parameter optimization
in this paper. \BLUE \respto{1M4}Prior to this paper, we focused extensively on transfer learning
and hyper-parameter optimization at prominent venues~\cite{krishna2018bellwethers,krishna2016too,agrawal2019dodge,agrawal2021simpler} because 
we believed those were ``state-of-the-art'' 
(and   had convinced 
many
journal reviewers of that
proposition). After this paper, our views have changed and like many other 
researchers, we believe we have
needless elaborated an 
inherently very simple process.\BLACK

\subsection{Literature Survey}\label{tion:ls}

\begin{table*}[!b]
\caption{50 `highly cited' (more than 10 citations per year) papers that ran one or more \textit{Cross} experiment(s) since 2011.}
\label{tbl:lit_review}
\centering
{\scriptsize

\begin{tabular}{r|l|r|r|r|r} 
\rowcolor[HTML]{EFEFEF} 
\textbf{Year}                    & \textbf{\#Data}                             & \textbf{\#Features} & \textbf{Projects} & \textbf{Cites/Year} & \textbf{Paper}                \\ \hline 
\textbf{2011}                    & \all                   & 20                 & 2                 & 16.44               & \cite{menzies2011local}       \\ \hline
                                 & \all                   & 198                & 6                 & 19.25               & \cite{li2012sample}           \\
                                 & \all                   & 17                 & 10                & 39.13               & \cite{ma2012transfer}         \\
                                 & \all                   & 20                 & 10                & 30.13               & \cite{he2012investigation}    \\
                                 & \all                   & 20                 & 7                 & 23.38               & \cite{menzies2012local}       \\
                                 & \all                   & 20                 & 2                 & 13.88               & \cite{bettenburg2012think}    \\
\multirow{-6}{*}{\textbf{2012}}  & \multirow{-7}{*}{\all} & 8                  & 9                 & 25.38               & \cite{rahman2012recalling}    \\ \hline 
                                 &                                            & 16                 & 41                & 14                  & \cite{turhan2013empirical}    \\
                                 &                                            & 17                 & 8                 & 48.43               & \cite{nam2013transfer}        \\
                                 &                                            & 20                 & 41                & 22.43               & \cite{peters2013better}       \\
                                 &                                            & 20                 & 10                & 19.57               & \cite{canfora2013multi}       \\
                                 & \multirow{-5}{*}{}                         & 20                 & 10                & 13.29               & \cite{peters2013balancing}    \\ 
                                 & \all                   & 20                 & 14                & 13.29               & \cite{herbold2013training}    \\
                                 & \all                   & 20                 & 10                & 11.43               & \cite{he2013learning}         \\
\multirow{-8}{*}{\textbf{2013}}  & \multirow{-3}{*}{\all} & 54                 & 12                & 32                  & \cite{rahman2013and}          \\ \hline 
                                 &                                            & 14                 & 11                & 16.5                & \cite{fukushima2014empirical} \\
                                 & \multirow{-2}{*}{}                         & 20                 & 10                & 20.5                & \cite{panichella2014cross}    \\
\multirow{-3}{*}{\textbf{2014}}  & \all                   & 26                 & 1398              & 18.5                & \cite{zhang2014towards}       \\ \hline 
                                 &                                            & 18                 & 7                 & 13.4                & \cite{ryu2015hybrid}          \\
                                 &                                            & 20                 & 11                & 19.8                & \cite{chen2015negative}       \\
                                 &                                            & 20                 & 10                & 14.4                & \cite{zhang2015empirical}     \\
                                 &                                            & 20                 & 17                & 12.4                & \cite{peters2015lace2}        \\
                                 &                                            & 20                 & 10                & 11.2                & \cite{canfora2015defect}      \\
                                 & \multirow{-6}{*}{}                         & 28                 & 11                & 24.4                & \cite{jing2015heterogeneous}  \\
\multirow{-7}{*}{\textbf{2015}}  & \all                   & 20                 & 10                & 33                  & \cite{he2015empirical}        \\ \hline 
                                 & None                                       & \textbf{X}                  & 26                & 35.5                & \cite{zhang2016cross}         \\
                                 &                                            & 14                 & 11                & 26.5                & \cite{kamei2016studying}      \\
                                 &                                            & 14                 & 6                 & 19.75               & \cite{yang2016effort}         \\
                                 &                                            & 17                 & 10                & 23.5                & \cite{ryu2016value}           \\
                                 &                                            & 20                 & 10                & 35.5                & \cite{xia2016hydra}           \\
                                 &                                            & 20                 & 21                & 20.75               & \cite{jing2016improved}       \\
                                 &                                            & 20                 & 30                & 11.75               & \cite{ryu2016effective}       \\
                                 &                                            & 26                 & 1390              & 16.25               & \cite{zhang2016towards}       \\
                                 & \multirow{-8}{*}{}                         & 61                 & 23                & 10.5                & \cite{krishna2016too}         \\
\multirow{-10}{*}{\textbf{2016}} & \all                   & \textbf{X}                  & 10                & 70.25               & \cite{wang2016automatically}  \\ \hline 
                                 &                                            & 20                 & 15                & 21.33               & \cite{ryu2017transfer}        \\
                                 &                                            & 61                 & 34                & 80                  & \cite{nam2017heterogeneous}   \\
\multirow{-3}{*}{\textbf{2017}}  &                                            & 61                 & 8                 & 10.67               & \cite{ni2017cluster}          \\ \hline 
                                 &                                            & 6                  & 58                & 21.5                & \cite{zhou2018far}            \\
                                 &                                            & 14                 & 6                 & 20.5                & \cite{chen2018multi}          \\
                                 &                                            & 20                 & 11                & 23                  & \cite{hosseini2018benchmark}  \\
                                 &                                            & 61                 & 18                & 18.5                & \cite{krishna2018bellwethers} \\
                                 &                                            & 61                 & 28                & 14.5                & \cite{li2018cost}             \\
                                 & \multirow{-9}{*}{}                         & 20                 & 10                & 14                  & \cite{wang2018deep}           \\
                                 & \all                   & 61                 & 16                & 19.5                & \cite{wu2018cross}            \\
\multirow{-8}{*}{\textbf{2018}}  & \multirow{-2}{*}{\all} & \textbf{X}                  & 10                & 20                  & \cite{dam2018deep}            \\ \hline 
                                 &                                            & 14                 & 6                 & 17                  & \cite{huang2019revisiting}    \\
                                 & \all                   & 61                 & 8                 & 14                  & \cite{chen2019multiview}      \\
                                 & \all                   & 20                 & 7                 & 20                  & \cite{chen2019software}       \\
\multirow{-4}{*}{\textbf{2019}}  & \multirow{-3}{*}{\all} & 20                 & 14                & 19                  & \cite{liu2019two}            \\ \hline 
\end{tabular}}
~\\~\\KEY: \ \fcolorbox{black}{all}{} Part, \fcolorbox{black}{white}{} All, \textbf{None} - No training data, \\\textbf{X} - Features automated
\end{table*}

\begin{table*}[!t]
\centering
\caption{44 `highly cited' (more than 10 citations per year) papers that built software defect prediction models applying methods based on ensemble/tuning/both in the past  decade (2011 to 2020).}
\label{tbl:lit_review_2}
{\scriptsize
\begin{tabular}{r|rlc|rr|r|rlc|rr}
\rowcolor[HTML]{C0C0C0} 
\textbf{Year} &
  \textbf{Cites/Year} &
  \textbf{Type} &
  \textbf{\#Data} &
  \textbf{Projects} &
  \textbf{Paper} &
  \textbf{Year} &
  \textbf{Cites/Year} &
  \textbf{Type} &
  \textbf{\#Data} &
  \textbf{Projects} &
  \textbf{Paper} \\ \cline{3-3} \cline{9-9} \hline
2011 &
  \multicolumn{1}{l|}{37.9} &
  \multicolumn{1}{l|}{} &
  All &
  2 &
  \cite{bird2011don} &
  2013 &
  \multicolumn{1}{l|}{25} &
  \multicolumn{1}{l|}{} &
  All &
  6 &
  \cite{jiang2013personalized} \\
2012 &
  \multicolumn{1}{l|}{17.33} &
  \multicolumn{1}{l|}{} &
  All &
  14 &
  \cite{sun2012using} &
  2014 &
  \multicolumn{1}{l|}{33} &
  \multicolumn{1}{l|}{} &
  All &
  6 &
  \cite{ye2014learning} \\
2013 &
  \multicolumn{1}{l|}{54.13} &
  \multicolumn{1}{l|}{} &
  All &
  10 &
  \cite{wang2013using} &
  2014 &
  \multicolumn{1}{l|}{36.86} &
  \multicolumn{1}{l|}{} &
  All &
  20 &
  \cite{scandariato2014predicting} \\
2013 &
  \multicolumn{1}{l|}{24} &
  \multicolumn{1}{l|}{} &
  All &
  41 &
  \cite{peters2013better} &
  2016 &
  \multicolumn{1}{l|}{15.8} &
  \multicolumn{1}{l|}{} &
  All &
  1,390 &
  \cite{zhang2016towards} \\
2014 &
  \multicolumn{1}{l|}{22} &
  \multicolumn{1}{l|}{} &
  All &
  10 &
  \cite{panichella2014cross} &
  2016 &
  \multicolumn{1}{l|}{33.4} &
  \multicolumn{1}{l|}{} &
  \all Part &
  17 &
  \cite{fu2016tuning} \\
2015 &
  \multicolumn{1}{l|}{53.33} &
  \multicolumn{1}{l|}{} &
  All &
  29 &
  \cite{ghotra2015revisiting} &
  2016 &
  \multicolumn{1}{l|}{79.8} &
  \multicolumn{1}{l|}{} &
  \all Part &
  10 &
  \cite{wang2016automatically} \\
2015 &
  \multicolumn{1}{l|}{13.33} &
  \multicolumn{1}{l|}{} &
  All &
  7 &
  \cite{ryu2015hybrid} &
  2017 &
  \multicolumn{1}{l|}{29.25} &
  \multicolumn{1}{l|}{} &
  All &
  6 &
  \cite{lam2017bug} \\
2015 &
  \multicolumn{1}{l|}{18.17} &
  \multicolumn{1}{l|}{} &
  All &
  6 &
  \cite{siers2015software} &
  2017 &
  \multicolumn{1}{l|}{47.25} &
  \multicolumn{1}{l|}{} &
  \all Part &
  7 &
  \cite{li2017software} \\
2016 &
  \multicolumn{1}{l|}{29.2} &
  \multicolumn{1}{l|}{} &
  All &
  11 &
  \cite{kamei2016studying} &
  2018 &
  \multicolumn{1}{l|}{37} &
  \multicolumn{1}{l|}{} &
  All &
  9 &
  \cite{agrawal2018better} \\
2016 &
  \multicolumn{1}{l|}{22.4} &
  \multicolumn{1}{l|}{} &
  All &
  21 &
  \cite{jing2016improved} &
  2018 &
  \multicolumn{1}{l|}{44} &
  \multicolumn{1}{l|}{} &
  All &
  8 &
  \cite{agrawal2018wrong} \\
2016 &
  \multicolumn{1}{l|}{15.8} &
  \multicolumn{1}{l|}{} &
  All &
  12 &
  \cite{wang2016multiple} &
  2019 &
  \multicolumn{1}{l|}{14} &
  \multicolumn{1}{l|}{} &
  All &
  10 &
  \cite{agrawal2019dodge} \\
2017 &
  \multicolumn{1}{l|}{31} &
  \multicolumn{1}{l|}{} &
  All &
  6 &
  \cite{yang2017tlel} &
  2019 &
  \multicolumn{1}{l|}{18.5} &
  \multicolumn{1}{l|}{} &
  \all Part &
  14 &
  \cite{liu2019two} \\
2017 &
  \multicolumn{1}{l|}{22.5} &
  \multicolumn{1}{l|}{} &
  All &
  15 &
  \cite{ryu2017transfer} &
  2020 &
  \multicolumn{1}{l|}{28} &
  \multicolumn{1}{l|}{\multirow{-13}{*}{TUNING}} &
  All &
  32 &
  \cite{kamei2016studying} \\ \cline{9-9}
2017 &
  \multicolumn{1}{l|}{16.25} &
  \multicolumn{1}{l|}{} &
  \all Part &
  11 &
  \cite{rathore2017towards} &
  2015 &
  \multicolumn{1}{l|}{32.67} &
  \multicolumn{1}{l|}{} &
  All &
  7 &
  \cite{tan2015online} \\
2018 &
  \multicolumn{1}{l|}{32.67} &
  \multicolumn{1}{l|}{} &
  All &
  27 &
  \cite{song2018comprehensive} &
  2015 &
  \multicolumn{1}{l|}{44} &
  \multicolumn{1}{l|}{} &
  All &
  6 &
  \cite{laradji2015software} \\
2018 &
  \multicolumn{1}{l|}{21.33} &
  \multicolumn{1}{l|}{} &
  All &
  15 &
  \cite{huda2018ensemble} &
  2016 &
  \multicolumn{1}{l|}{37.2} &
  \multicolumn{1}{l|}{} &
  All &
  10 &
  \cite{xia2016hydra} \\
2018 &
  \multicolumn{1}{l|}{27.67} &
  \multicolumn{1}{l|}{} &
  All &
  12 &
  \cite{tong2018software} &
  2016 &
  \multicolumn{1}{l|}{50.8} &
  \multicolumn{1}{l|}{} &
  All &
  18 &
  \cite{tantithamthavorn2016automated} \\
2019 &
  \multicolumn{1}{l|}{26} &
  \multicolumn{1}{l|}{} &
  All &
  44 &
  \cite{xu2019software} &
  2016 &
  \multicolumn{1}{l|}{24.4} &
  \multicolumn{1}{l|}{} &
  All &
  10 &
  \cite{ryu2016value} \\
2019 &
  \multicolumn{1}{l|}{21.5} &
  \multicolumn{1}{l|}{} &
  \all Part &
  10 &
  \cite{pascarella2019fine} &
  2018 &
  \multicolumn{1}{l|}{31} &
  \multicolumn{1}{l|}{} &
  All &
  18 &
  \cite{bowes2018software} \\
2020 &
  \multicolumn{1}{l|}{28} &
  \multicolumn{1}{l|}{} &
  All &
  2 &
  \cite{rhmann2020software} &
  2018 &
  \multicolumn{1}{l|}{19.67} &
  \multicolumn{1}{l|}{} &
  All &
  6 &
  \cite{wang2018deep} \\
2020 &
  \multicolumn{1}{l|}{31} &
  \multicolumn{1}{l|}{} &
  All &
  3 &
  \cite{chekam2020selecting} &
  2018 &
  \multicolumn{1}{l|}{47.67} &
  \multicolumn{1}{l|}{} &
  \all Part &
  18 &
  \cite{tantithamthavorn2018impact} \\
2020 &
  \multicolumn{1}{l|}{20} &
  \multicolumn{1}{l|}{\multirow{-22}{*}{ENSEMBLE}} &
  All &
  12 &
  \cite{pandey2020bpdet} &
  2019 &
  \multicolumn{1}{l|}{13.5} &
  \multicolumn{1}{l|}{\multirow{-9}{*}{BOTH}} &
  All &
  25 &
  \cite{zhou2019improving} \\ \cline{3-3} \cline{9-9} \hline
\end{tabular}%
}
~\\~\\KEY: \ \fcolorbox{black}{all}{} \all Part, \fcolorbox{black}{white}{} All
\end{table*}

To understand more about  late-data (data-hungry) reasoning in SE, we queried Google Scholar~\footnote{https://scholar.google.co.in/ in 2020} to find software defect prediction articles (similar to our prior work~\cite{icse21}) in three areas:
\bi

\item We found 982 articles in Google Scholar using the query (``cross-project defect prediction'')
in the last ten years.

\item We found more than 1,740  articles in the last five years alone  (searching the query  ``defect prediction'' AND ``(tuning OR hyper-parameter optimization) and 3,050  articles using the query ``defect prediction'' AND ``ensemble''. 
\ei

Following the advice of Agrawal et al.~\cite{agrawal2018better}, and Mathews et al.~\cite{mathew2018finding} we focused
only on ``highly cited'' papers, i.e., those with more than ten citations per year.
Reading those  papers, and after discarding papers pure of survey nature, we filtered the papers that performed some transfer learning experiments or a complex method ( ensemble or tuning).  We summarized the results of this literature survey in  \tbl{lit_review} (Transfer learning) and \tbl{lit_review_2} (Tuning and Ensemble methods).

Within those three sets, we found
three approaches to row
selection:
\bi
\item `All' if the methods of that paper used all \respto{3F}\BLUE rows (training data instances) from one or more projects to \BLACK build defect predictors. 
\item `Part' if the methods of that paper explore a
large search space of one or more projects to find a small set of rows that are worthy of transfer data.  
\item In one case, in 2016, we also found a `None' approach
that used no training but just clustered the test data to find outliers, which were then labeled as bugs~\cite{zhang2016cross}. Having recorded that
method, this paper will not explore this minority approach.
\ei
Note that regardless of being ``All'' or ``Part'', the analysis looks at the
data across the entire life cycle before returning some or all of it.

As to other kinds of data
selection, for all these papers,
we counted:
\bi

\item The number of projects  used in those studies;
\item The number of features used by their predictors.
\item The applied any ensemble or tuning or both techniques.
\ei

When we tried to place 
this paper into \tbl{lit_review} and \tbl{lit_review_2}, we found
that our approach was, quite
literally, off the charts.
 The methods advocated by this paper are
neither ``whole'' nor ``part'' since we learn from early data, then stop collecting (so unlike all the research in \tbl{lit_review} and \tbl{lit_review_2}, we never look
at all the data). Further, our ``number of projects''=1 (which does not
even appear in \tbl{lit_review})
and we only use a handful of features (far fewer than the features used by other work in
\tbl{lit_review}).

Hence we assert, with some confidence,
that the methods of this paper have not been
previously explored.

\subsubsection{Representative Techniques}
In order to design an appropriate experiment, we used the following guidelines.

Firstly, we are comparing the early sample to sampling over
a larger space of project data. Hence, in the following,
we will show {\em early} versus {\em all} experiments.

Secondly, there are two ways to find data: 
{\em within-project} and {\em cross-project}. 
Therefore, we will divide the {\em early} early-data
and {\em all} late-data
experiments into:
\bi
\item early-data: early-within and early-cross
\item late-data: all-within and all-cross
\item Complex Methods: Optimizers (DODGE and HyperOpt) and ensemble method (TLEL).
\ei

Lastly, looking into the literature, we can see some
clear state-of-the-art algorithms that should be represented
in our study (specifically, the $TCA+$ and $Bellwether$ cross-project learning methods~\cite{krishna2016too,nam2013transfer}). Accordingly, when exploring {\em cross-project} learning, we will employ those
methods.

Explaining the exact methods used in this study requires
some further details on those algorithms. Please see 
\tion{learners} for the algorithm details and for a discussion of what algorithms we
selected.
But, in summary, to the best of our knowledge, the early-data variants of these techniques have not been explored before in SE.

\begin{table*}[!b]
\caption{14 Commit level features that Commit Guru tool \cite{kamei2012large,rosen2015commit} mines from GitHub repositories}
\label{tbl:features}
\scriptsize
\begin{tabular}{|l|l|l|}
\hline
\rowcolor[HTML]{C0C0C0} 
\textbf{Dimension}                    & \textbf{Feature} & \textbf{Definition}                                           \\ \hline
                                      & NS               & Number of modified subsystems \respto{2J}\BLUE(subsystem identified using root directory (package) name) \BLACK                                \\ 
                                      & ND               & Number of modified directories                                \\ 
                                      & NF               & Number of modified files                                      \\ 
\multirow{-4}{*}{\textbf{Diffusion}}  & ENTROPY          & Distribution of modified code across each file                \\ \hline
                                      & LA               & Lines of code added                                           \\ 
                                      & LD               & Lines of code deleted                                         \\ 
\multirow{-3}{*}{\textbf{Size}}       & LT               & Lines of code in a file before the change                            \\ \hline
\textbf{Purpose} & FIX & Whether or not this change is a defect fixing change. \\ \hline
                                      & NDEV             & \#developers  changing   modified files          \\ 
                                      & AGE              & Mean time   from   last to the current change \\ 
\multirow{-3}{*}{\textbf{History}}    & NUC              & \#changes to   modified files before         \\ \hline
                                      & EXP              & Developer experience                                          \\ 
                                      & REXP             & Recent developer experience                                   \\ 
\multirow{-3}{*}{\textbf{Experience}} & SEXP             & Developer experience on a subsystem                           \\ \hline
\end{tabular}
\end{table*}

 \begin{figure*}
\begin{subfigure}{0.43\textwidth}
\includegraphics[height=1.6in]{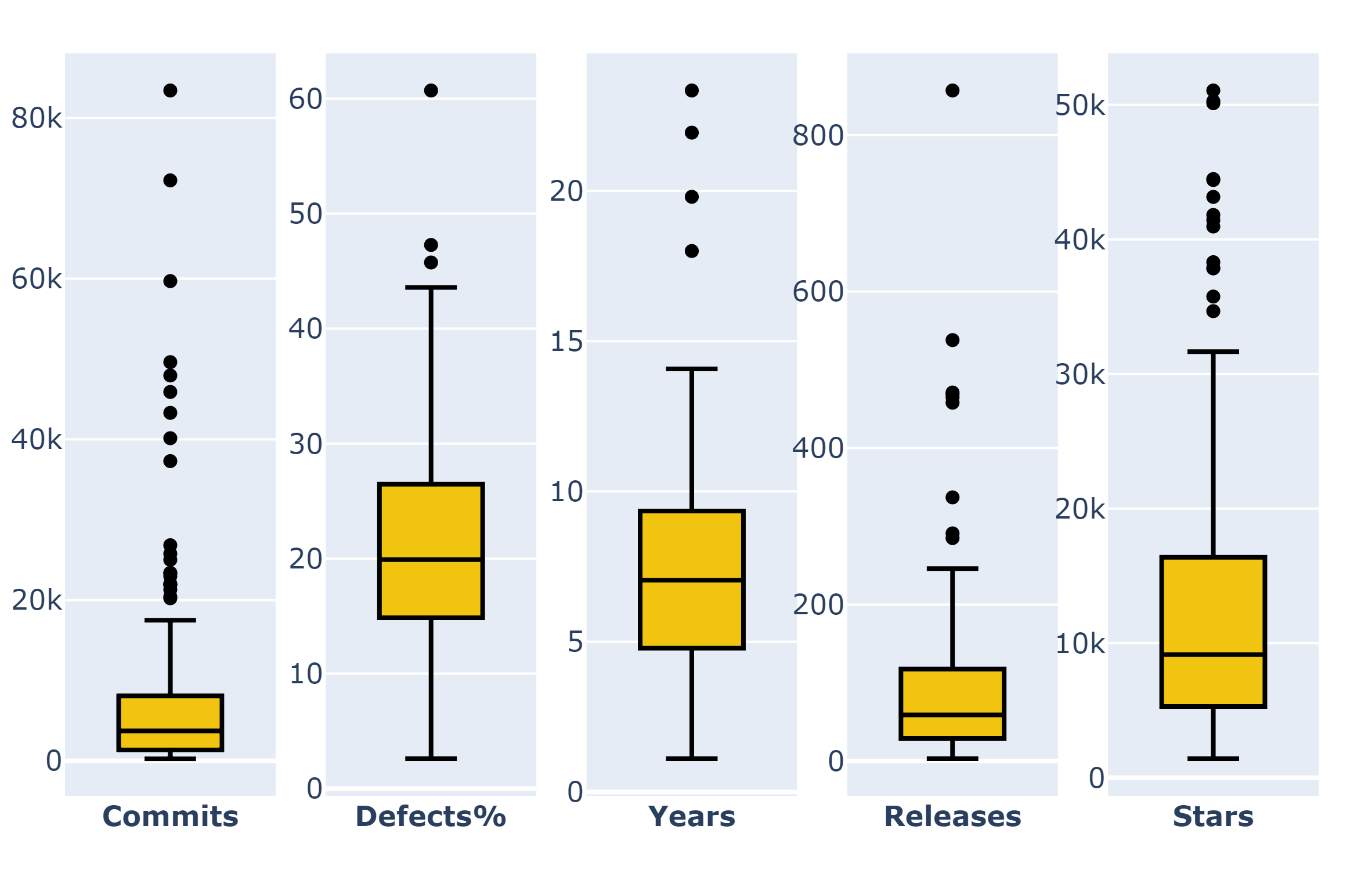}  
 \caption{\textbf{Popular projects} (stars $\mathit{>1000}$).\\
\textit{Distributions seen in all 1.2 millions of commits of all \\155 \textit{popular} projects: median values of commits \\(3,728), percent of defective commits (20\%), life \\span in years (7), releases (59) and stars (9,149).}}\label{fig:distribution_popular}~\\
\end{subfigure}~~~~~~~~~\begin{subfigure}{0.43\textwidth}

\vspace{-5mm}\includegraphics[height=1.475in]{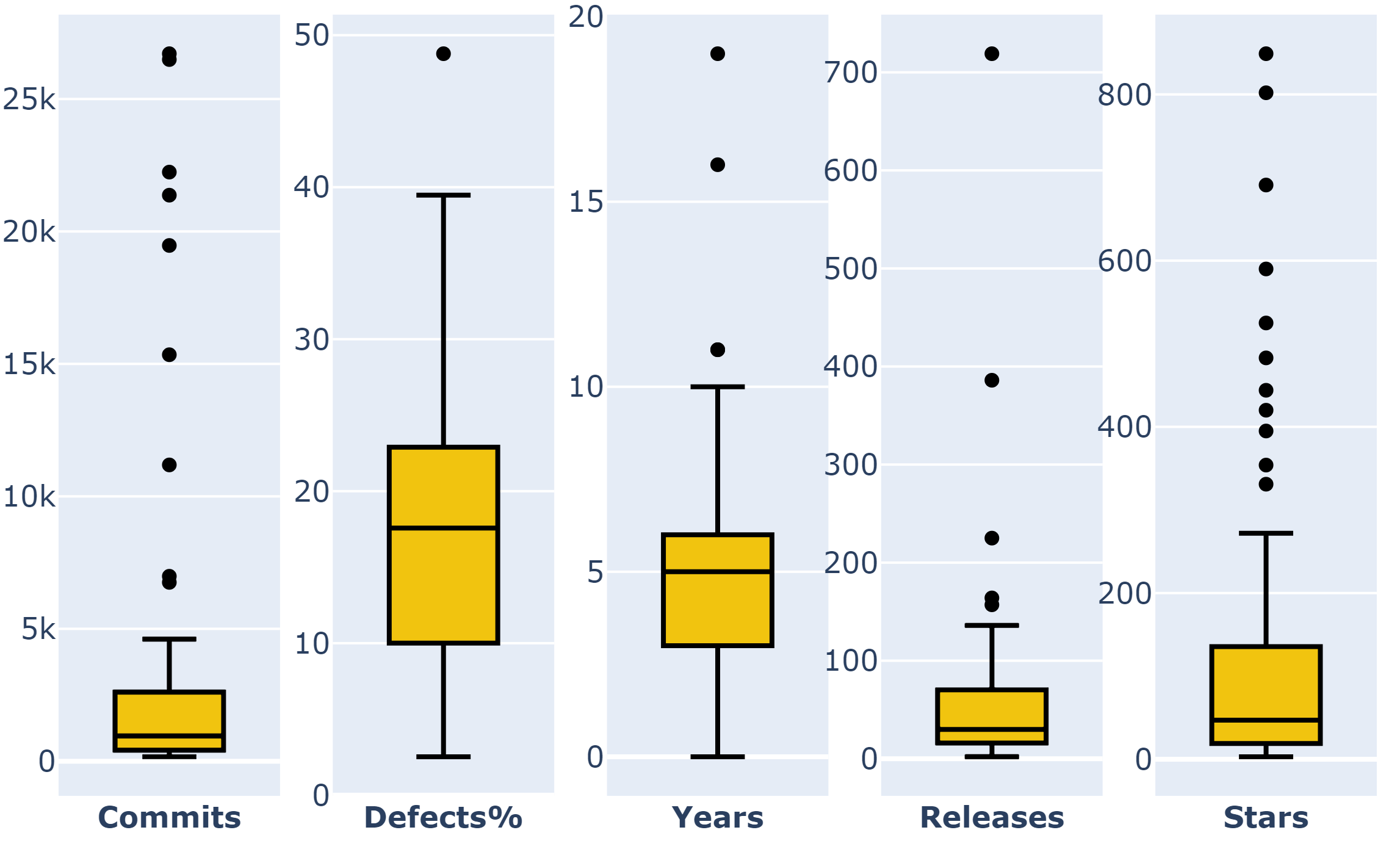} 
\caption{\textbf{Unpopular projects} (stars $\mathit{<1000}$). \\\textit{Distributions seen in all 258,000+ commits of all 89 \textit{unpopular} projects: median values of commits (957), percent of defective commits (18\%), 
life span in years (5), releases (30) and stars (47).}}\label{fig:distribution_unpopular}

\end{subfigure}

\caption{240 GitHub project data distributions.}\label{fig:distribution}

\end{figure*}

%
\begin{figure*}
\begin{subfigure}{0.45\textwidth}
\includegraphics[width=2.5in,keepaspectratio]{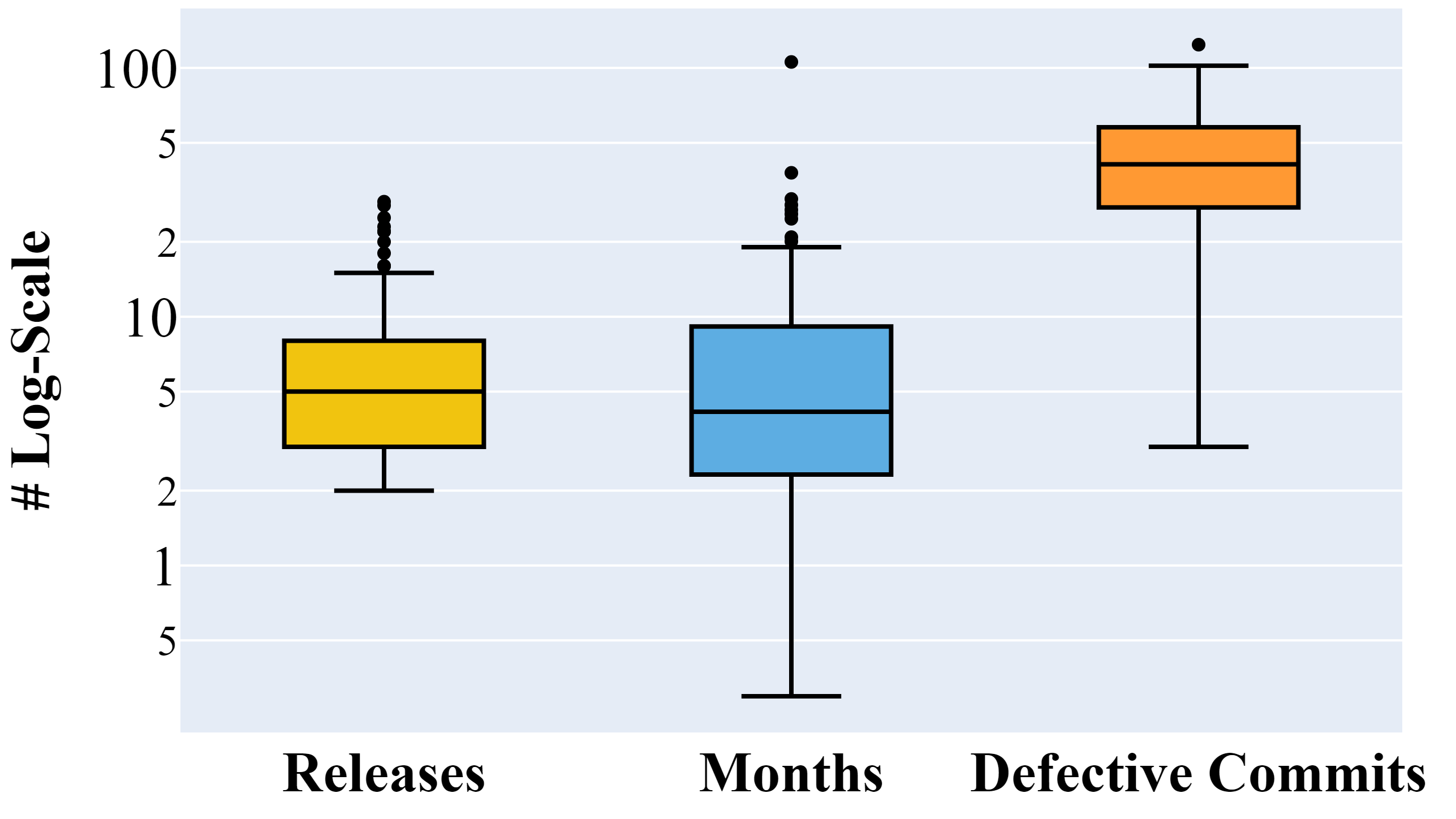} 
 \caption{Popular projects. Median= 5 releases, 4 months,  and 41 defective commits.}\label{fig:commits_popular}
\end{subfigure}
~~~~~\begin{subfigure}{0.45\textwidth}
\includegraphics[width=2.5in,keepaspectratio]{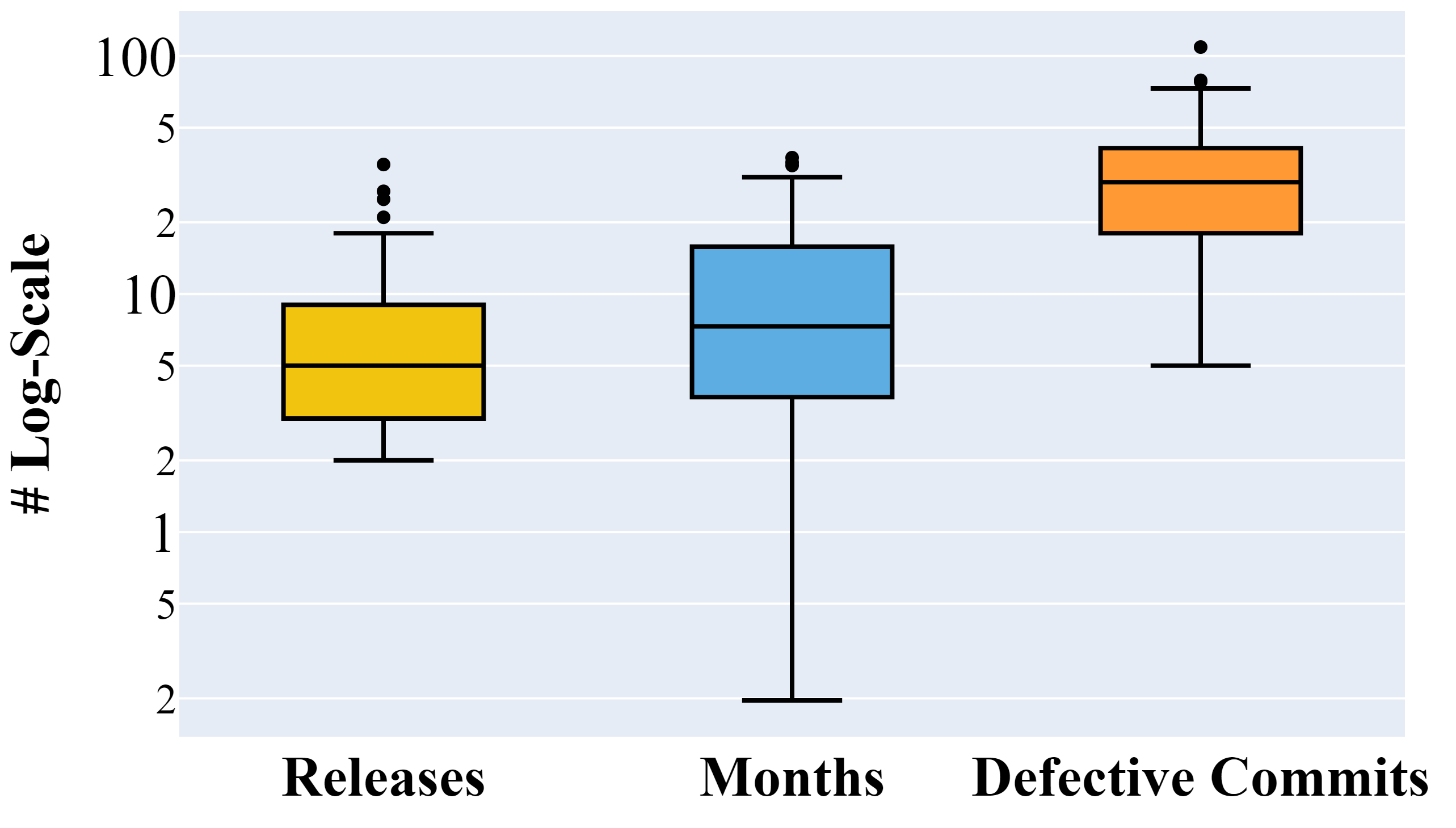} 
 \caption{Unpopular projects. Median= 5 releases, 7 months, and  29 defective commits.}\label{fig:commits_unpopular}
\end{subfigure}
\caption{Distributions  seen in the first 150 commits. }
\label{fig:commits_distribution}
\end{figure*}

We understand the above areas do not cover all of the software defect prediction literature but to the best of our knowledge, we assert that we covered a range of active research areas within defect prediction to test the scope of our conclusion about early methods.

\section{Experimental Methods}\label{tion:methods}

The rest of this paper assesses supervised defect prediction models built using the early-bird heuristic versus other learning policies and optimization techniques for the data of Figure~\ref{fig:whale}. 

At a high level, we build a defect prediction model in three steps. 

First, we sample training commits using one of the sampling methods listed in \tion{sampling_methods}. The test data is the commits from a project release. In the case of within-project defect prediction, past commits are never used as test commits. In all the experiments of this paper, there is no overlap of commit data between test and training data.  Second, training and test commits are pre-processed using the techniques listed in \tion{pre}. Third, we build the model using one of the classifiers listed in \tion{learners}.  

We use the measures listed in \tion{measures} to assess the performance of the defect prediction model's ability to classify test commits (as defective or clean). This process of training, testing, and evaluation is repeated for every applicable project release in all 240 projects. Apart from RQ1, all other RQs consider all 240 GitHub projects for analysis. 

Lastly, to compare the overall performance of various treatments we apply the Scott-Knott test discussed in \tion{sk}. Here treatment  is a defect prediction model built using a specific `sampling method', `classifier' and  evaluated on a particular `measure.' For example, one treatment can be a logistic regression-based defect prediction model that uses an early-bird (early-data) sampling method. Another could be a decision-tree-based model that used a data-hungry (late-data) sampling method. The population in this example is the distribution of `Recall' scores evaluated in all the applicable project releases.  The two treatments are then ranked (clustered) using the Scott-Knott test on the basis of an evaluation measure (say `Recall').

\subsection{Data} \label{tion:data}

All data used in this study comes from open-source projects hosted on GitHub that other SE researchers can replicate. 
For details on that data, see Table~\ref{tbl:features}, Figure~\ref{fig:distribution} and Figure~\ref{fig:commits_distribution}.

 \respto{2F}\BLUE We reuse the 155 popular projects from our prior study which was randomly sampled from the project list curated by Munaiah et al.~\cite{munaiah2017curating}. Additionally, we collected 85 unpopular projects to answer RQ1 using the same process as our prior work, except that we are looking for projects with less than $1000$ stars. In both popular and unpopular projects collected, we started with sampling 100s of projects from the Munaiah et al. list but only 155 popular projects and 85 unpopular projects satisfied the non-trivial and sanity checks.

To repeat from the prior work,  we rely on the criteria listed by Munaiah et al. to differentiate an engineering project from a trivial one.  Then we used Commit-Guru's public portal to \respto{1M5} mine these project data (commits) along with their fourteen process metrics~\cite{rosen2015commit}. \BLUE \respto{2K}Similar to the back-trace approach by SZZ  algorithm~\cite{cite_szz_original} each commit was labeled ``defective'' (based on  certain defect-related keywords)  or ``clean'' (otherwise) internally by \textit{Commit Guru}.  \textit{Commit-guru} walks back into the code to find the changes associated with that commit.  \BLACK Projects were rejected according to the  standard sanity checks listed in prior work~\cite{icse21,yan2020just}: 

\bi
\item Less than 1\% defective commits;
\item Less than two releases;
\item Less than one year of activity;
\item No license information;
\item Less than five defective and five clean commits.
\ei


\respto{1E} \BLUE Further, looking at Figure~\ref{fig:distribution},
we see that above and below 1000 stars, the distributions are 
different (for example, the median number of commits is 3,728 and 957). \BLACK
 
The projects collected in this way were developed in widely-used programming languages (including  Java, Python, C, C++, C\#, Kotlin, JavaScript, Ruby, PHP, Fortran, Go, Shell, etc.) for various domains.  

All 14  features used in this study are listed in \tbl{features}. Those features are extracted from these projects using  \textit{Commit Guru}~\cite{rosen2015commit}. The use of these particular features has been prevalent and endorsed by prior studies ~\cite{kamei2012large,rahman2013and}. \textit{Commit Guru} publicly available tool used in numerous works~\cite{xia2016predicting,kondo2020impact} based on a 2015 ESEC/FSE paper. 
Those 14 features became the independent attributes used in this empirical study.

In light of results by Nagappan and Ball, we created relative churn and standardized LA (lines of code added) and LD (lines of code deleted) features by separating by LT (lines of code in a file before the change) and LT and NUC (the number of unique changes to the modified files
before) dividing by NF (number of modified files)~\cite{nagappan2005use,kondo2019impact}. Likewise, we dropped ND (number of modified directories) and REXP (recent developer experience) since Kamei et al. revealed that NF and ND are correlated with REXP and EXP (developer experience). Lastly, we applied the logarithmic transformation to the remaining features (except for the boolean variable 'FIX') to handle skewness~\cite{shihab2010understanding}.

\begin{figure}[!t]
\begin{center}
\end{center}
\end{figure}
 
 This empirical study uses three sets of algorithms:
 \bi
 
 \item
 The nine classification algorithms described in \tion{learners};
 \item
Pre-processing algorithms (for some sampling policies)
described in \tion{pre};
\item
 The seven sampling methods are described in \tion{sampling_methods}.
 \ei

But, \textit{CommitGuru} does not provide  project release information. Therefore we cloned each GitHub project to our local machine and extracted GitHub releases/tags information by executing the following command shown below:

\begin{verbatim}
git log --tags --simplify-by-decoration --pretty="format:%ai %d"
\end{verbatim}

\subsection{Classifiers}\label{tion:learners}

We use all the six classifiers and optimizers used in our prior work~\cite{icse21}. Those six classifiers are prevalent in the SE literature chosen from the \respto{2L} \BLUE four groups \BLACK tabulated by Ghotra et al. ~\cite{ghotra2015revisiting}. 
The classifiers we include in this empirical study are:
\bi
\item Logistic Regression (LR);
\item Nearest neighbor  (KNN) (minimum 5 neighbors);
\item Decision Tree (DT);
\item  Random Forrest (RF)
\item Na\"ive Bayes (NB);
\item Support Vector Machines (SVM)
\ei

\subsection{Complex Methods}
This section describes the optimizers and an ensemble learning technique that is prevalent for software defect prediction. They are used in augmenting classifiers to yield better predictive performance.

\subsubsection{\textbf{DODGE:}}\label{tion:dodge} Agrawal et al.'s DODGE is a state-of-the-art hyper-parameter optimization method extensively assessed for software defect prediction~\cite{agrawal2019dodge}.  One critical problem in hyper-parameter optimization (such as grid search or brute force) is the run-time overhead. DODGE overcomes this by terminating much faster by skipping redundant options.

\begin{figure}[!tbp]
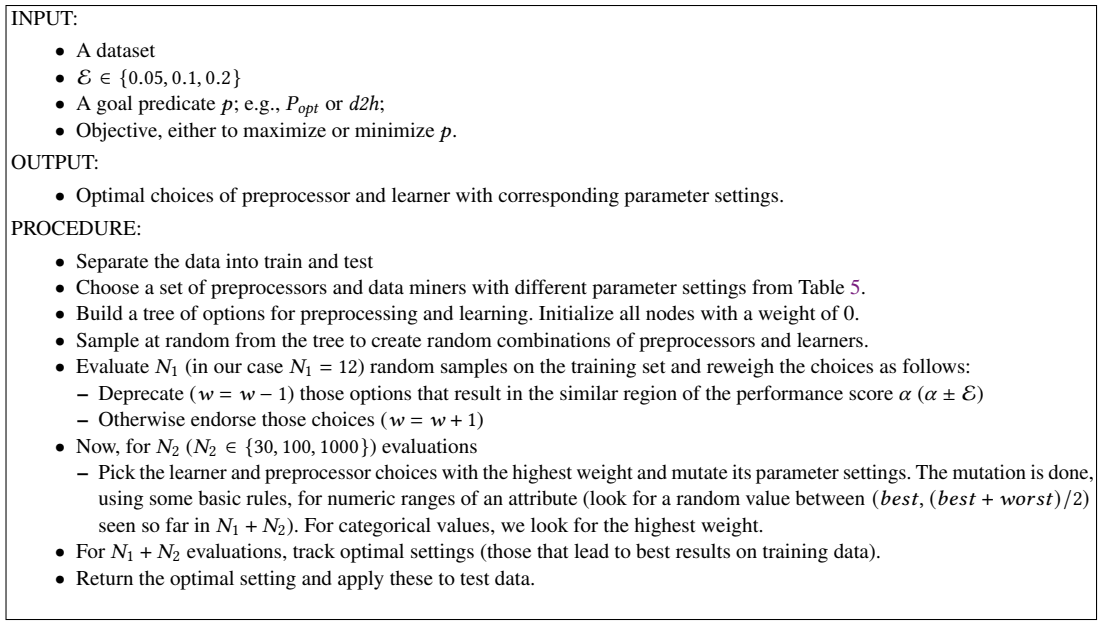

\captionsetup{justification=centering}
\small \begin{tabular}{|p{.95\linewidth}|}\hline
INPUT: 
\bi
\item A dataset
\item $\mathcal{E} \in \{0.05, 0.1, 0.2\}$
\item A goal predicate $p$; e.g., ${P_{\mathit{opt}}}$ or  $\mathit{d2h}$;
\item Objective, either to maximize or minimize $p$.
\ei
OUTPUT:
\bi\item Optimal choices of preprocessor and learner with   corresponding parameter settings.
\ei 
PROCEDURE:
\bi
\item Separate the data into train and test
\item Choose a set of preprocessors and data miners with different parameter settings from Table~\ref{tbl:dodge}.
\item Build a tree of options for preprocessing and learning.   Initialize all   nodes with a weight of 0.
\item Sample at random from the tree to create  random combinations   of preprocessors and learners. 
\item Evaluate $N_1$ (in our case $N_1=12$) random samples  on the training set and reweigh the choices as follows:
    \bi
    \item   Deprecate ($w=w-1$)  those options that result in the similar region of the performance score $\alpha$ ($\alpha \pm\mathcal{E}$)
    \item Otherwise endorse those choices ($w=w+1$)
    \ei
        
\item Now, for $N_2$ ($ N_2 \in \{30, 100, 1000\}$) evaluations 
    \bi
    \item Pick the learner and preprocessor choices with the  highest weight and mutate its parameter settings. The mutation is done, using some basic rules, for numeric ranges of an attribute (look for a random value between $(best, (best+worst)/2)$ seen so far in  $N_1+ N_2$). For categorical values, we look for the highest  weight.
    \ei
\item For $N_1+ N_2$ evaluations,  track     optimal settings (those that lead to  best results on training data).
\item Return the optimal setting and apply these to test data.
\ei

\\\hline
\end{tabular}
\caption{Pseudo-code of DODGE replicated from~\cite{agrawal2019dodge}}
\label{fig:dodge_pcode}
\end{figure}

DODGE is an ensemble tree of classifiers and pre-processors as shown in \tbl{dodge}. DODGE is broadly a two-step process as shown in \fig{dodge_pcode}. First, DODGE iteratively shrinks (prunes the tree) the tuning search space by ignoring redundant options sampled from the tree. Then in the next set of iterations, it finds near-optimal options by looking between the best and worst options seen so far. 

DODGE is shown to perform much better than building models with classifiers or pre-processors directly with off-the-shelf default options~\cite{agrawal2019dodge}. Notably, Agrawal et al. have highlighted that DODGE fails on complex data sets. \respto{1F}\BLUE Here the complexity of a dataset is determined using Levina et al. intrinsic dimensionality ($\mu_D$) computation~\cite{levina2005maximum}. They say that many data sets that are stored in high-dimensional formats can actually be compressed without losing much information~\cite{Aggarwal01} and that Principal Component Analysis based methods frequently overestimate intrinsic dimensions.

Calculating the number of items found at distances within radius \textit{r} (where \textit{r} is the distance between two configurations) while varying \textit{r} yields the intrinsic dimension of a dataset with \textit{N} items. The intrinsic dimensionality is measured by this because:

\bi
\item If the items are distributed only in one $r=1$ dimension, we will only discover a linear increase in the number of items as $r$ rises.
\item If the items are dispersed across, say, $r>1$ dimensions, we will discover polynomially more items as $r$ grows.
\ei

As shown in Equation~\ref{eq:cr}, 
Levina et al. normalize the number of items based on the number of $N$ items that are being compared. They advise reporting the number of intrinsic dimensions ($\mu_D$) as the maximum value of the slope between  $\mathit{ln(r)}$ vs  $\mathit{ln}(C(r))$. The  $\mathit{ln}(C(r))$ value computed as follows:

\begin{equation}\label{eq:cr}
C(r) = \frac{2}{N(N-1)} \sum_{i=1}^N \sum_{j=i + 1}^N I (||x_i,x_j||<r) 
\end{equation}

Note that equation~\ref{eq:cr} above uses the L1-norm rather than the Euclidean L2-norm to calculate  because Courtney et al.~\cite{Aggarwal01} suggest that for data with many columns,   L1 performs better than L2. Accordingly, DODGE yields the best performance for data sets with low dimensionality ($\mu_D\approx 3$) and below par performance for higher-dimensional data ($\mu_D > 8$). Notably, the intrinsic dimensionality of 240 projects explored in this study is $\mu_D < 2$ (compatible with DODGE to perform). \BLACK

\begin{table}

\scriptsize
\caption{Hyperparameter tuning options explored in this paper (taken from~\cite{agrawal2019dodge}). Note that we make no claim that this is a complete list of options. Rather, we merely claim that a reader of the recent SE literature on hyperparameter optimization might be tempted
to try some subset of the following.}
\label{tbl:dodge}
\begin{tabular}{|p{.95\linewidth}|}\hline
\textbf{DATA PRE-PROCESSING}

\noindent
Software defect prediction:

\bi
\item \textbf{Transformations}
\bi
\item StandardScaler
\item MinMaxScaler
\item MaxAbsScaler
\item RobustScaler(quantile\_range=(a, b))
    \bi
    \item a,b= randint(0,50), randint(51,100)
    \ei
\item KernelCenterer
\item QuantileTransformer(n\_quantiles=a,\newline output\_distribution=c, subsample=b)
    \bi
    \item a, b = randint(100, 1000), randint(1000, 1e5)
    \item c=randchoice([`normal',`uniform'])
    \ei
\item Normalizer(norm=a)
    \bi
    \item a = randchoice([`l1', `l2',`max'])
    \ei
\item Binarizer(threshold=a)
    \bi
    \item a= randuniform(0,100)
    \ei
\ei
\ei

\\\hline
\textbf{LEARNERS}

\noindent
Software defect prediction and text  mining:
\bi
\item DecisionTreeClassifier(criterion=b,\newline splitter=c, min\_samples\_split=a)
    \bi
   \item a= randuniform(0.0,1.0)
   \item b, c= randchoice([`gini',`entropy']),\newline randchoice([`best',`random'])
   \ei
\item RandomForestClassifier(n\_estimators=a,criterion=b,  min\_samples\_split=c)
    \bi
   \item a,b = randint(50, 150), randchoice(['gini', 'entropy'])
   \item c = randuniform(0.0, 1.0)
   \ei
\item LogisticRegression(penalty=a, tol=b, C=float(c))
    \bi
    \item a=randchoice([`l1',`l2'])
    \item b,c = randuniform(0.0,0.1), randint(1,500)
    \ei 
\item MultinomialNB(alpha=a)
    \bi
    \item a= randuniform(0.0,0.1)
    \ei 
\item KNeighborsClassifier(n\_neighbors=a, weights=b, p=d,\newline metric=c)
    \bi
    \item a, b = randint(2, 25), randchoice([`uniform', `distance'])
    \item c = randchoice([`minkowski',`chebyshev'])
    \item if c=='minkowski': d= randint(1,15)  else:  d=2
    \ei
\ei  
\\\hline
\end{tabular}
\end{table}

\subsubsection{\textbf{Hyperopt:}} A prevalent hyper-parameter optimizer proposed by Bergstra et al. in ~\cite{Bergstra:2011}. Agrawal et al., in a  recent software defect prediction work~\cite{agrawal2021simpler} showed DODGE to outperform hyperopt for software defect prediction. However, we include Hyperopt because it is a state-of-the-art optimizer in the machine learning community that may augment early methods and work differently compared to DODGE. Formally, hyperopt is a framework that can work with different optimization algorithms. For our experiments, we used Sequential model-based optimization and Tree-structured Parzen Estimator (TPE) wrapped in the Hyperopt toolkit \respto{1M6} \BLUE available online~\cite{bergstra2013making}. \BLACK

Hyperopt stochastically inputs hyper-parameter settings to TPE. The TPE groups the evaluations of various hyper-parameter settings to best and rest. Essentially, TPE reflects on the history of evaluations seen up to date to jump to the next best setting to explore.  The TPE explores the best hyper-parameter setting from the history of evaluations seen to date by order. \respto{1M7} \BLUE It then selects the best hyper-parameter options and is then modeled as a Gaussian with its own mean and standard deviation. \BLACK

\subsubsection{\textbf{Two-layer ensemble learning (TLEL):}}\label{tion:tlel} 
\begin{figure*}[!t]
\begin{center}
\includegraphics[width=4in]{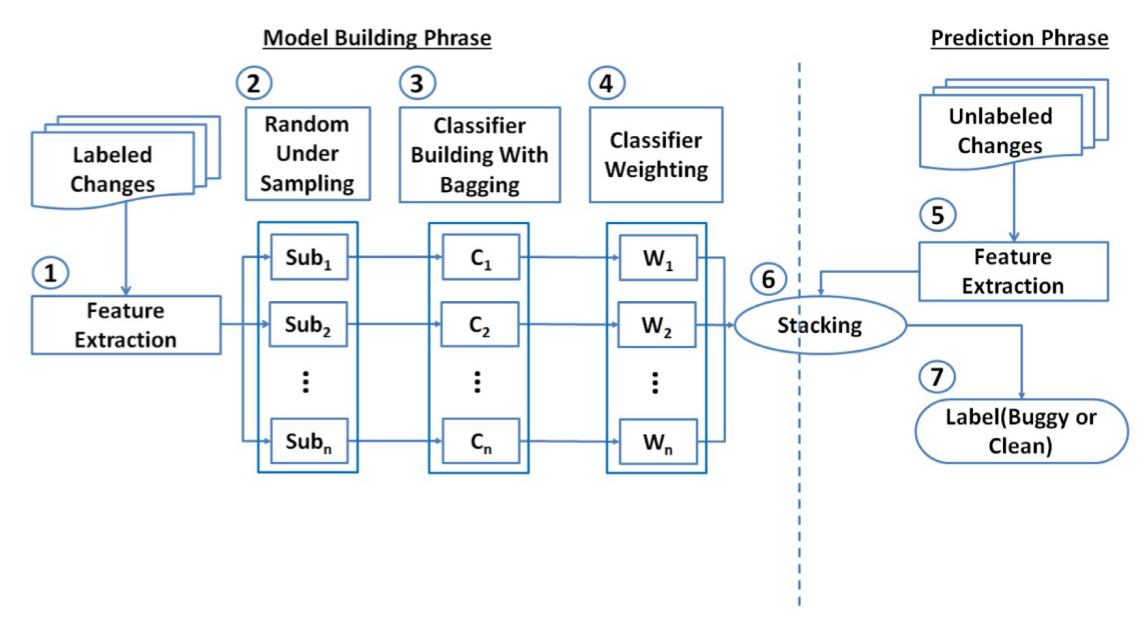}
\end{center}
\caption{Two-Layer Ensemble Learning framework replicated from the work by Yang et al.~\cite{yang2017tlel}  }\label{fig:tlel} 
\end{figure*}

Yang et al. in 2017 proposed an ensemble approach that leverages \respto{1M8} \BLUE decision trees \BLACK with bagging similar to a Random Forest model~\cite{yang2017tlel} depicted in \fig{tlel}. A difference here is that Random Forest uses many decision trees, but in TLEL, the training data is under-sampled randomly to train many \textit{yet} different Random Forest models. A commit is classified by TLEL as defective when most of the stacked and trained random Forest models agree (second layer ensemble).

\subsection{\textbf{No Data Methods:}} In the past, Menzies et al. showed  trivial approaches that use no training information like Manual Up/Down outperformed in identifying defects compared to complex methods~\cite{menzies2010defect}. Recently (2018) Zhou et al. reported that simple size-based models show a promising predictive performance, therefore advising researchers to include Manual Up/Down methods as a baseline while proposing any new technique~\cite{zhou2018far}.

Koru et al. related a module's defect proneness to its size. In one case with two commercial projects,  they found smaller modules were defective~\cite{koru2008theory} whereas in another they found larger classes were more defect prone~\cite{koru2008investigation}. Therefore we include both their  methods \textit{ManualDown} and \textit{ManualUp} to \respto{1G}\BLUE label \BLACK our test commits.

\subsection{ManualDown} We \respto{1G}\BLUE label \BLACK all test commits as defective if the size (`la' in \tbl{features}) is greater than the median size among the test commits~\cite{zhou2018far}. The philosophy here is more enormous changes should be inspected first and therefore penalized.

\subsection{ManualUp} We \respto{1G}\BLUE label \BLACK all test commits as defective if the size (`la' in \tbl{features}) is less than or equal to the median size among the test commits~\cite{koru2008investigation}. The philosophy here is more minor changes should be inspected first and therefore penalized.


\begin{table*}[!t]

\scriptsize
\begin{tabular}{c|c|c|p{2.25cm}|p{2.25cm}|p{3cm}}
\hline
\rowcolor[HTML]{EFEFEF} 
\textbf{Nature} & \textbf{Type} & \textbf{Method}               & \textbf{Pre-processing}                                                               & \textbf{\# Features (columns)}                                                 & \textbf{\# of commits (rows)}                                                                                          \\ \hline

      &           & Bellwether~\cite{krishna2016too}                    & CFS, SMOTE and steps illustrated in \tion{data} & Selected by  CFS.                                                      & All commits from the identified cross-project.   \\ 
late-data     & \textit{Cross}          & TCA+\cite{nam2013transfer}                          & SMOTE      & 5 components with a linear kernel (data supplied with all features)   & Pick the first 150 and last 150 commits from the bellwether project.                                                                    \\ \hline

              &            & \early   
$*\mathit{E_{size} (Bellwether)}$ & Steps illustrated in \tion{data} in~\cite{nagappan2005use,kamei2012large,kondo2020impact}  & LA, LT = lines added, lines of code in the file before change                                                          & Sample an equal number of defective and clean commits as available in the first 150 commits (not exceeding 25 each). \\ 

early-data                  & \textit{Cross}          & \early $*\mathit{E_{size} (TCA+)}$         & Steps illustrated in \tion{data}                & 2 components with the linear kernel (data supplied only with LA and LT) & \\ \hline

  &            & \early   $\mathit{E}$~\cite{icse21} & CFS and steps illustrated in \tion{data}        & Selected by  CFS                                                        &  \\ 

early-data                   & \textit{Within}          & \early $*   \mathit{  E_{size}}$   & Steps illustrated in \tion{data}                & LA and LT                                                           &  (same as above) \\  \hline
late-data                   & \textit{Within}          & $ALL$  & CFS, SMOTE and steps illustrated in \tion{data} & Selected by  CFS.   &  All the commits before the release, which is under test. \\

\end{tabular}

~\\~\\KEY:  All early-data sampling methods are shaded (\fcolorbox{early}{early}{}) and policies proposed in this study are indicated by *. 
\caption{This table lists three baseline approaches (where two of them are \textit{Cross} and the remaining one is a \textit{Within}) along with three early-data variants to support this study. Note, many of these methods used the steps advised in~\cite{nagappan2005use,kamei2012large,kondo2020impact}. Those steps are discussed in \tion{data}. Further, for a discussion on CFS, see \tion{pre}.}
\label{tbl:policies}
\end{table*}

\subsection{Pre-processers}\label{tion:pre}

Pre-processing the training and in some cases, test data has been shown very effective and necessary in building defect predictors. The following methods are representative of \BLUE \respto{1M9}\respto{3M6} \respto{2M}  much of \BLACK
of the prior research in this field:
\bi
\item The steps advised by \cite{nagappan2005use,shihab2010understanding,kondo2019impact}
(see  \tion{data});
\item Correlation-based Feature Selection (CFS);
\item Synthetic Minority Over-Sampling (SMOTE).
\ei
Note: Not all combinations of pre-processing steps are valid.
\tbl{policies} show the types of pre-processing applied for each sampling policy used in this paper. Further, we avoid a common threat to validity, we state there that  we applied SMOTE only to the training data but {\em not} to the test data~\cite{agrawal2018better}).

 

\subsubsection{Correlation-based Feature Selection (CFS):} CFS is a prevalent feature selection technique proposed by Hall ~\cite{hall2003benchmarking}. That is often used in building supervised software defect prediction models, especially while building defect predictors~\cite{kondo2019impact}. A heuristic-based technique to incrementally assess a subset of features. Internally, a best-first search is applied to identify influential sets of features that are not correlated with each other. Nevertheless, it should be correlated with the target (classification). Each feature subset is computed as follows:
 $\mathit{merit}s = \respto{1M10} \BLUE k  * r{\mathit{cf}}/\sqrt{k+k(k-1)r_{\mathit{ff}}}$ \BLACK
where:
\bi
\item 
$\mathit{merit}s$ is the value of  subset $s$ having $k$ features; 
\item
$r{\mathit{cf}}$ is the score that explains the connection of that feature set to the class;
\item
$r{\mathit{ff}}$ is the feature to feature mean and  connection between the items in $s$.
\ei 
 
\subsubsection{Synthetic Minority Over-Sampling (SMOTE):} A sampling technique proposed by Chawla et al.~\cite{chawla2002smote} to handle the class imbalance problem. Because when there are an unequal number of clean and defective commits (or modules, records, etc.) in the training set, classifiers can scuffle to discover the target class. SMOTE is recommended by many researchers in the space of software defect prediction~\cite{agrawal2018better,tantithamthavorn2018impact}. 


\respto{1H} \BLUE \subsection{Training Data Selection  Methods}\label{tion:sampling_methods}
\begin{wrapfigure}{l}{2.2in}
\begin{center}
   \includegraphics[width=2.2in,keepaspectratio]{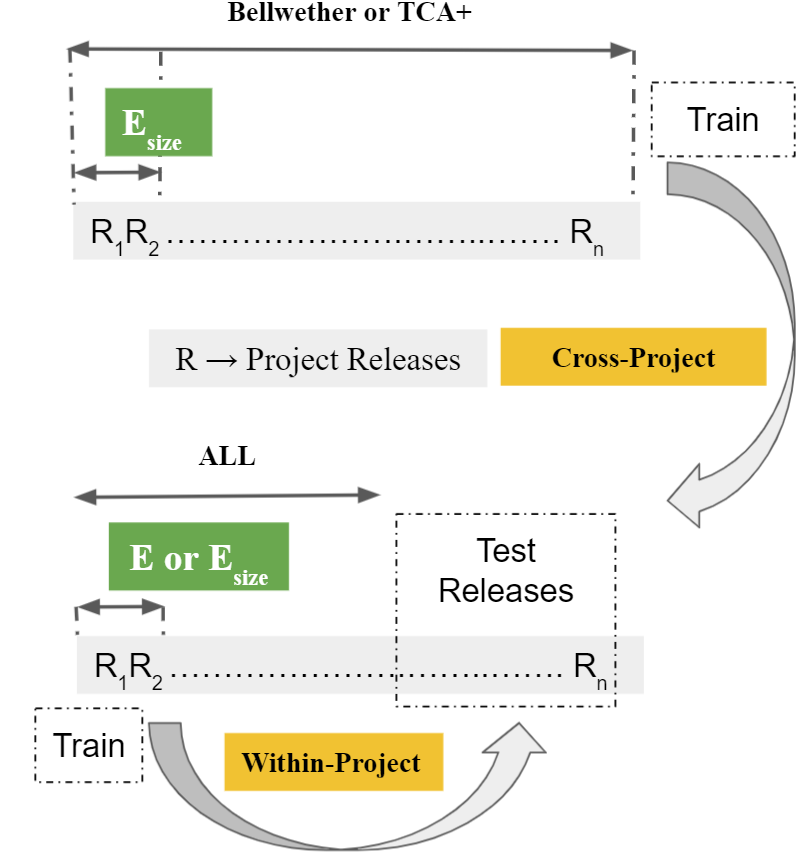}
\caption{A visual map to gauge the sampling policies (training commits selection) listed in \tbl{policies}.
}\label{fig:policies}  
\end{center}
\end{wrapfigure}

 The essence of this work is how we sample data to train defect predictors and figure~\ref{fig:policies}  attempts to portray the same. The figure~\ref{fig:policies} shows a set of options from which we can generate a large number of sampling options.
For example, training data come from {\em within} or be drawn {\em cross} from another project.  It is helpful to view figure~\ref{fig:policies} along with the selection methods listed in \tbl{policies}.


Shrikanth et al. identified many sampling strategies (based on recent releases, recent months of data, and all past data) to build predictors in the \textit{Within} project context~\cite{icse21}. To find representative training data selection methods for the \textit{Cross} context we checked the papers listed in \tbl{lit_review}. We found numerous \textit{Cross} techniques reported in the past decades; for example, Sousuke's work recently (2020) explored 24 \textit{Cross}  approaches~\cite{amasaki2020cross}. Overall, in the context of \textit{Cross}, we found only two (`whole' or `part'). Nevertheless, both  `whole' and `part' is late-data.  As the focus of this study is to check the efficacy of using early-data approaches in prevalent \textit{Cross} approaches and not rank numerous \textit{Cross} techniques.

\subsubsection{Cross:}
We chose two representative \textit{Cross} techniques, specifically $\mathit{TCA+}$ and $\mathit{Bellwether}$, because:

\bi
\item $\mathit{TCA+}$ is an active \textit{Cross} technique in this space based on the seminal work by Nam et al.~\cite{nam2013transfer}.
\item $\mathit{Bellwether}$, a recent (2018) baseline approach by Krishna et al. that performed better than $\mathit{TCA+}$~\cite{krishna2018bellwethers}
\ei

\subparagraph{\textbf{Bellwether Project ($ \mathit{Bellwether}$): }} \label{tion:bellwether} 
The bellwether transfer learning method assumes ``\textit{When a community works on software, then there exists one exemplary project, called the bellwether, which can define predictors for the others.}~\cite{krishna2018bellwethers}''.


 \subparagraph{\textbf{Transfer Component Analysis (TCA+):}} \label{tion:tcap} Pan et al. proposed a domain adaptation technique that enables the transition of information between source and target domains~\cite{pan2010domain}. Extending that, Nam et al. stacked many normalization rules on top of TCA to form TCA+~\cite{Nam13}. A specific normalization rule is applied based on the similarity of the data set characteristics between the source and the target project. Rahul and Menzies~\cite{krishna2018bellwethers} report TCA+ to perform better than other transfer methods, namely, Transfer Naive Bayes~\cite{ma2012transfer} and Value Cognitive Boosting Learner~\cite{ryu2016value}.  
TCA+ can be somewhat computationally expensive.
To ally that,  we built defect predictions with varying sizes in our prior work~\cite{icse21} and found that predictors needed a minimum of 150 training commits. As mentioned above unlike other methods $TCA+$ is CPU intensive, therefore we could only increase training commits up to 300 (2x 150). This is not a sampling threat because later in \tion{results} we find the early-data-lite $TCA+$ variant performs better than $TCA+$ trained with 300 commits supporting our early conjecture.
For TCA+ related policies listed in \tbl{policies} we reused the implementation from their replication package~\footnote{https://sailhome.cs.queensu.ca/replication/featred-vs-featsel-defectpred/} by Kondo et al., which is an EMSE'19 article about feature reduction techniques on software defect prediction models~\cite{kondo2019impact}. The above two techniques use all available cross-project data. We include two Early \textit{Cross} variants specifically Early Bellwether ($\mathit{E_{size} (Bellwether)}$), Early TCA+ ($\mathit{E_{size} (TCA+)}$) that only uses early cross-project data and only two features.

\subsubsection{Within:}
$ALL$ uses all available past data within the project under test to predict future defects. We create two early variants ($\mathit{E}$)  that use early within project data will all features (as selected by CFS) and ($\mathit{E_{size}}$) is same as $E$ except it uses only two features 

For specific sampling numbers and rules please refer to \tbl{policies}.   \BLACK

\subsection{Evaluation Criteria}\label{tion:measures}

We use all eight criteria used in the baseline study to gauge the predictors built using various sampling policies. That study consults from widely-used measures~\cite{
menzies2008implications,%
wang2013using,%
tantithamthavorn2018impact,%
bennin2019relative,%
zhang2014towards,%
mcintosh2017fix,%
8494821,%
kondo2020impact,%
yatish2019mining,%
yan2019characterizing,%
zhang2016use,%
d2012evaluating,%
kamei2012large} in the software defect prediction literature.


\subsubsection{Brier}
\respto{1J}\BLUE
The brier score is  the mean squared error (MSE) of the actual outcome  $y \in \{0,1\}$(clean or defective) and the predicted probability estimate  $p = \operatorname{Pr}(y = 1)$ for test commits (samples) computed as below:

\begin{equation}
\footnotesize
Brier = \frac{1}{n_{\text{samples}}} \sum_{i=0}^{n_{\text{samples}} - 1}(y_i - p_i)^2
\end{equation}

Numerous software defect prediction papers~\cite{mcintosh2017fix,tantithamthavorn2018impact,8494821,kondo2020impact} endorse this measure~\cite{pedregosa2011scikit}.
 \BLACK

\subsubsection{Initial number of False Alarms (IFA)} 

Based on the observation by Parnin and Orso ~\cite{parnin2011automated}  that developers lose their trust in such analytics if they encounter many initial false alarms. Thus by simply counting the number of false alarms encountered after sorting the commits in the order of probability of being defect-prone. IFA is simply the number of false alarms before finding the first actual alarm.

\subsubsection{Recall} Recall is the number of inspected defective commits divided by all the defective commits.

\begin{equation}
\footnotesize
\mathit{Recall} = \frac{\mathit{True\ Positives}}{\mathit{True\ Positives + False\ Negatives}} 
\end{equation}

\subsubsection{False Positive Rate (PF)} 

PF is the ratio between the number of clean commits predicted as defective to all the defective commits (irrespective of classification). 

\begin{equation}
\footnotesize
\mathit{PF} = \frac{\mathit{False\ Positives}}{\mathit{False\ Positives + True\ Negatives}}
\end{equation}

\subsubsection{Area Under the Receiver Operating Characteristic curve (AUC)} It is simply the area under the curve between the false-positive rate and true-positive rate.

\subsubsection{Distance to Heaven (D2H)} D2H or ``distance to heaven'' is computed as an aggregation on two metrics Recall and False Positive Rate (PF). Where ``heaven'' is a place with $\mathit{Recall = 1}$ \&
$\mathit{PF = 0}$~\cite{chen2018applications}.

\begin{equation}
\footnotesize
\mathit{D2H}  = \frac{\sqrt{(1-\mathit{Recall})^2 + (0-\mathit{PF   })^2}}{ \sqrt{2}}
\end{equation}

\subsubsection{G-measure (GM)} 
GM is computed as a harmonic mean between the complement of PF and Recall. It is measured as shown below:
\begin{equation}
\footnotesize
\mathit{G-Measure}  = \frac{2 * \mathit{Recall} * (1-\mathit{PF})}{ \mathit{Recall}+(1-\mathit{PF})}
\end{equation}

GM and D2H essentially combine the same two measures, $\mathit{Recall}$ and $\mathit{PF}$. Nevertheless, we still employ those as they have been used and endorsed separately in the literature. Notably, as seen from results in~\cite{icse21} and in this work shown later in \tion{results}, it is not necessary that achieving good results on GM would also associate with good D2H (or vice-versa).

Due to the nature of the classification process, some criteria will  always offer contradictory results:
 \bi
 \item
 A classifier may simply achieve 100\% $\mathit{Recall}$ just by labeling all the test commits as defective.
 But as a side-effect, that method will incur a high $\mathit{PF}$.
\item
 Secondly, a classifier may classify all test commits as clean to show 0\% $\mathit{PF}$, but that method will incur a very low $\mathit{Recall}$.
 \item
 
 Lastly, Brier and Recall are also antithetical since reducing the loss function implies missing some conclusions lowering $\mathit{Recall}$.
 \ei

\subsubsection{Mathew's Correlation Coefficient (MCC)}\label{mcc} MCC  utilizes all the four computations namely True Positives (TP), True Negatives (TN), False Positives (FP),  and False Negatives (FN) of the confusion matrix, such that:
\begin{equation}
\footnotesize
{\text{MCC}}=\frac {{\mathit {TP}}\times {\mathit {TN}}-{\mathit {FP}}\times {\mathit {FN}}}{\sqrt {({\mathit {TP}}+{\mathit {FP}})({\mathit {TP}}+{\mathit {FN}})({\mathit {TN}}+{\mathit {FP}})({\mathit {TN}}+{\mathit {FN}})}}
\end{equation}

Thus, many researchers endorse the use of MCC, especially in the space of software defect prediction~\cite{yao2020assessing,kondo2020impact}. It returns a score between $-1$ and $+1$, where $+1$ indicates higher predictive performance, $-1$ contrary predictions, and $0$ indicates most of the predictions' poor predictive performance (random).

Note for the above eight predictive performance measures:
\bi
\item Initial number of False Alarms range from 0 to \# (maximum number of commits inspected);
\item MCC range from -1 to 1;
\respto{1I} \BLUE \item {\em D2H, IFA, Brier, PF} of these criteria need to be minimized, i.e., for these criteria {\em less} is {\em better}.
\item For    four of these {\em AUC, Recall, G-Measure, and MCC} criteria need to be maximized, i.e., for these criteria {\em more} is {\em better}\BLACK.
\ei 

Prior work has shown that precision has significant issues for unbalanced data. We do not include that in our evaluation ~\cite{menzies2008implications}.  Prior reviewers of this paper have noted that this might mean we miss certain effects relating to that section of the evaluation space. To address that point, we have added an evaluation metric that draws from multiple ``corners'' of the evaluation space (see the MCC measure of \S\ref{mcc})

\subsection{Statistical Tests}\label{tion:sk}


Predictors built using each of the sampling policies listed in \tbl{policies} are tested on all appropriate (future) project releases. Then we compute each of the evaluation criteria discussed above in \tion{measures}. Then each of the scores is grouped and exported by a sampling policy and classifier pair. 

To reiterate, each population is a collection of a specific evaluation score. Moreover, populations can have the same median but have an entirely different distribution. Thus to rank each of those populations of evaluation scores, we use the Scott-Knott test recommended by Mittas et al. in TSE'13 paper~\cite{Mittas13}.  This procedure is a top-down bi-clustering method used to rank different predictors created using different sampling policies and classifiers under a specific evaluation measure. This method sorts a list of $l$ those evaluation scores $\mathit{ls}$ by their median score. Next, it then splits $l$ into sub-lists \textit{m, n}. This is done to maximize the expected value of differences in the observed performances before and after divisions. 

For lists $l,m,n$ of size $\mathit{ls},\mathit{ms},\mathit{ns}$ where $l=m\cup n$, the ``best'' division maximizes $E (\Delta)$; i.e.
the difference in the expected mean value
before and after the spit: 

\[
E (\Delta)=\frac{ms}{ls}abs (m.\mu - l.\mu)^2 + \frac{ns}{ls}abs (n.\mu - l.\mu)^2
\]

Additionally, a conjunction of bootstrapping and A12 effect size test by Vargha and Delaney~\cite{vargha2000critique} is applied to avoid ``small effects'' with statistically significant results.  

\textbf{Important note:} we apply the Scott-Knott test 
independently to all the eight evaluation criteria; i.e., when we compute ranks, we do so for (say) $\mathit{Recall}$ separately to $\mathit{PF}$.  \respto{1A} \BLUE In the rest of this paper, when we say `\textit{similar}', we mean the distributions in comparison belong to the same Scott-Knott rank (ie., clustered in the same group), and when we say `\textit{better}' we mean the distribution is statistically different (ie., cannot be clustered into the same group) and favorable (eg., higher $Recall$ or lower $PF$). \BLACK

\subsection{Other Details}\label{tion:other}

We wish to report not just performance scores but also offer some details about the model that generates those scores. To
that end, we will exploit the bellwether effect. Specifically, we will  (a)~find the most important features, then (b)~ use them in bellwethers that represent most of our data.

\section{Results} \label{tion:results}

An overview of our experiments is shown in  \fig{policies}. We build defect predictors using the sampling strategies listed in \tbl{policies} and classifiers elucidated in \tion{learners}. Then, similar to our prior work~\cite{icse21} we test the built defect predictors in all the project releases~\footnote{Exception: RQ1 only considers unpopular projects.}.

In this paper, the results of RQ1 are used in RQ2 and so on to eliminate redundant assessment and comparison of treatments. Therefore, we eliminate treatments (defect predictors) that are either already been assessed or when they do not affect the conclusion of the RQ under investigation.  We do this to eliminate redundancy in analysis and improve brevity in our results. For instance, we do not build defect predictors using training commits based on recent data  (since it was assessed in prior work against early-bird heuristic~\cite{icse21}). 

\subsection{\textbf{RQ1: Can we build early software defect prediction models from unpopular projects?}}

\respto{2G} \BLUE \subsubsection{Motivation} Our prior results that endorsed early methods were scoped to only popular GitHub projects~\cite{icse21}. Choosing only popular projects is a sampling decision often made by many SE researchers in empirical studies to mitigate generalizability threat~\cite{munaiah2017curating}. As mentioned earlier in \tion{data}, unpopular projects could be non-trivial projects (like homework assignments), and that could affect the conclusion. 

On the contrary, a strong argument could be that perhaps popular projects may not realistically represent SE in the real world. In other words, not all projects are fully staffed with a sufficient budget. Therefore it is necessary and worthy to check the value of early methods on the unpopular sample of projects. \respto{1M11} \BLUE To reiterate, in  \tion{data} we elucidate the selection criterion of unpopular and non-trivial SE projects from GitHub. Later in  \tion{threats} we discuss how we mitigated GitHub projects sampling threats.\BLACK

\subsubsection{Approach}
To check whether the proposed early method  work on unpopular projects we compare defect predictors built by sampling training commits using $E$ with those that sampled all past data ie., $ALL$. We do not consider other stratification like the release, or 3 months since $ALL$ subsumes more data. Notably, $ALL$ is a prevalent (50\%) sampling methods in software defect prediction~\cite{icse21}. Therefore to assess defect predictors on each project release we construct the experiment as follows:

\bi
\item First we sample training data within the project. The sampling will depend upon either $E$ or $ALL$. 
\item The sampled training data is pre-processed and appropriate features are  selected as listed in \tbl{features}.
\item Classifiers listed in \tion{learners} are instantiated with copies of pre-processed training data.
\item All the instantiated predictors are tested on all 85 unpopular project releases and their predictive performance is gauged using 7 measures elucidated in \tion{measures}.
\item Lastly, their scores (eg., the population of recall scores) per classifier+policy (pair) is ranked using the Scott-Knott test elucidated in \tion{sk}. 

\ei

To avoid a methodological error, we avoid overlap of train and test commits. For instance, we do not test on project releases before the first 150 commits. Notably, the population of evaluation measures (e.g, $Recall$) is tested on an equal number of unpopular project releases.

\subsubsection{Results}

To reiterate, we wanted to check if our early-bird effect identified in our prior work also holds among unpopular GitHub projects~\cite{icse21}. The results shown in \tbl{rq1} support our early-bird heuristic even among unpopular GitHub projects because the policy $E$ uses fewer early data (row \#1) and performs just as same as predictors that were built sampled with $ALL$ (all past commits) in row \#3. 

\begin{table*}[h]
\caption{ `12' \textit{within-project}  software defect prediction models tested `only' in 85 \textit{unpopular} project's releases.
 In the first row, ``+'' and ``-'' denote the criteria that need to be maximized or minimized, respectively. 
 All the other  rows show combinations of sampling policies
 and classifiers.
Green cells denote ``early-data'' sampling was employed. Cells marked in 
gray all have the same top rank as the
best results (and those
ranks were determined by the Scott-Knott algorithm
described in \S\ref{tion:sk}).
The `wins' column (see column \#3) counts how often a particular sampling policy/classifier achieves
a top score. Inter-quartile ranges are indicated within `( )'. }
\label{tbl:rq1}
\small
\begin{tabular}{llr|r|r|r|r|r|r|r|r}
\rowcolor[HTML]{EFEFEF} 
\textbf{Policy} &
  \textbf{Classifier} &
  \textbf{Wins} &
  \textbf{Recall+} &
  \textbf{PF-} &
  \textbf{AUC+} &
  \textbf{D2H-} &
  \textbf{Brier-} &
  \textbf{G-Score+} &
  \textbf{IFA-} &
  \textbf{MCC+} \\ \hline  
\early E & LR & \multirow{5}{*}{6} & \hil 75 (50) & 40 (34) & \hil 64 (17) & \hil 41 (22) & 38 (22) & \hil 65 (27) & \hil 1 (4) & \hil 24 (28) \\
\early E & SVM &  & \hil 69 (50) & 38 (30) & \hil 64 (18) & \hil 41 (21) & 37 (21) & \hil 63 (30) & \hil 1 (4) & \hil 23 (31) \\
ALL & SVM &  & 50 (54) & \hil 15 (18) & \hil 65 (25) & \hil 40 (32) & \hil 23 (14) & 54 (51) & \hil 1 (5) & \hil 28 (45) \\
ALL & RF &  & 50 (52) & \hil 16 (22) & \hil 64 (22) & \hil 42 (28) & \hil 25 (16) & 53 (46) & \hil 1 (4) & \hil 28 (41) \\
ALL & KNN &  & 50 (43) & \hil 20 (17) & \hil 64 (21) & \hil 40 (25) & \hil 26 (14) & 54 (41) & \hil 1 (4) & \hil 26 (38) \\\hline
\early E & KNN & 5 & \hil 67 (40) & 42 (26) & \hil 62 (19) & \hil 42 (19) & 40 (17) & \hil 61 (28) & \hil 2 (5) & 19 (30) \\ \hline
\early E & RF & 4 & 56 (52) & 28 (28) & \hil 63 (20) & \hil 42 (23) & 33 (18) & 56 (39) & \hil 2 (5) & \hil 23 (33) \\ \hline
ALL & NB & 2 & 14 (60) & \hil 9 (27) & 50 (13) & 64 (24) & \hil 25 (22) & 14 (51) & 4 (12) & 1 (30) \\ \hline
\early E & NB & \multirow{4}{*}{1} & 56 (66) & 35 (38) & 56 (19) & 48 (25) & 38 (24) & 52 (47) & \hil 2 (6) & 14 (32) \\ 
ALL & LR &  & 50 (80) & 25 (28) & 60 (19) & 47 (37) & 30 (20) & 51 (69) & \hil 2 (7) & 17 (36) \\
\early E & DT &  & 50 (53) & 32 (31) & 57 (19) & 47 (27) & 37 (23) & 52 (46) & \hil 2 (6) & 14 (33) \\
ALL & DT &  & 50 (38) & 30 (23) & 57 (18) & 45 (23) & 34 (18) & 51 (35) & \hil 2 (5) & 15 (34)
\end{tabular}%
\end{table*}


\begin{tcolorbox}[colback=red!5,colframe=gray!50]{\textit{Early methods are `not' scoped to only popular projects.}}
\end{tcolorbox}

\subsection{\textbf{RQ2: Can we build early software defect prediction models with fewer features?}}

\subsubsection{Motivation} The prior work showed that it is possible to build operable defect predictors using fewer early project data. One reason for the preceding work was that much of the knowledge required (defects) to build defect predictors were reported within a few months of the project. In this work, we explore that region again to understand which of those 14 features listed in \tbl{features} are essential to the early predictor $E$. And if the answer to that is a few, can we restrict the predictors to only learning from a fixed set of features while producing operable performance? It would be easier to explain with fewer features why a specific commit was classified as defective by the predictor.

\subsubsection{Approach} To answer this RQ we perform three experiments as follows:

\bi
\item In the first experiment we report the frequency of features chosen by CFS while building defect predictors which were sampled using $E$.

\item In the second experiment we create a new sampling policy (say $E_{?}$) that only uses the most frequently chosen feature identified in the above step.

\item We create two predictors that sampled training commits (pre-processed as listed in \tbl{policies}) using $E$ and $E_{?}$ and test on all project releases. 

\item In the third experiment, we also test all project releases using the ManualUp and ManualDown approach (elucidated in \tion{learners}) that uses no training information.  This is because researchers have advised including such trivial  approaches that do not use any training information~\cite{zhou2018far}.

\item Lastly, we compare the predictive performance of defect predictors that were built using $E$, $E_{?}$ and $ManualDown$ and $ManualUp$ on all eight evaluation measures listed in \tion{measures} and rank them using the Scott-Knott test elucidated in \tion{sk}.

\ei

\subsubsection{Results}

\begin{figure}[!t]
\begin{center}
\includegraphics[width=3.5in]{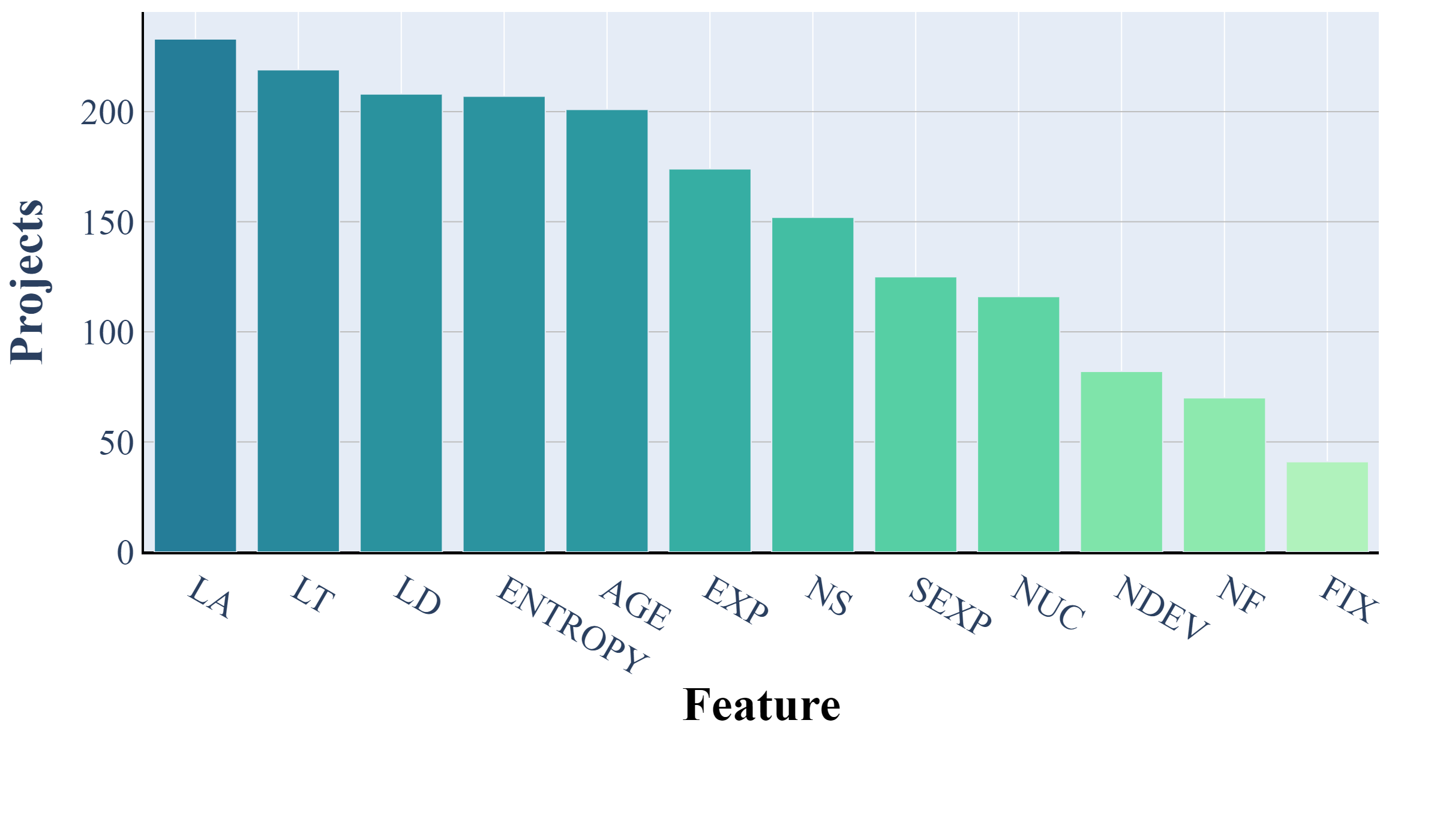}
\end{center}
\caption{Frequency of features chosen by the CFS feature selector (using just the first 150 commits) sampled using  $\mathit{E}$ in all the 240 projects}\label{fig:2_features} 
\end{figure}

We make the following observations from \fig{2_features} and \tbl{rq2}:

\subparagraph{Finding Important Features:} Shrikanth et al.'s early sampling rule `E' is shown to be adequate for \textit{Within}~\cite{icse21}.
Given that we are building models from such a sample,
it is tempting to ask, ``does that small sample
produce a succinct model?''. Note that if this was indeed the case, then we could offer a clear report on what factors most influence defects.

\fig{2_features} shows the frequency of features selected 
by the CFS algorithm (see \tion{pre}) across all our 240 projects  (using just the first 150 commits). To select the ``best''
features from that space, we build models using the top
$1\le x \le 12$  ranked features, stopping with
$x+1$ features performed  no \textit{better} than $x$ features (and here, we are
 testing all the releases that occurred
after that first   150 commits).
That procedure reported that models learned from
two size-ranked features performed as well as anything else:
\begin{itemize}
\item
LA: Lines of code added
\item
LT: Lines of code in a file before the change
\end{itemize}

Using the results from \fig{2_features} we built an early defect predictor only using the top two features `LA' and `LT'. Therefore $E_{?}$ becomes $E_{size}$. The results from \tbl{rq2} clearly show that $E_{size}$ have more wins than predictors that sampled training commit using $E$. And by extension to prior result $ALL$ or other stratification like recent release or based on months.  While $ManualDown$ is seen at the top of the table due to higher recall, it also produced many false alarms (see row \#3 where $PF$ is 48\%).

\begin{table*}[!t]
\caption{ `14' \textit{within-project}  software defect prediction models tested in all 240 project releases.
 In the first row, ``+'' and ``-'' denote the criteria that need to be maximized or minimized, respectively. 
 All the other  rows show combinations of sampling policies
 and classifiers.
Green cells denote ``early-data'' sampling was employed. Cells marked in 
gray all have the same top rank as the
best results (and those
ranks were determined by the Scott-Knott algorithm
described in \S\ref{tion:sk}). The score highlighted in bold  \textcolor{red}{\textbf{red}} text indicates undesirable PF, despite many wins on other evaluation measures.
The
`wins' column (see column \#3) counts how often a particular sampling policy/classifier achieves
a top score. Inter-quartile ranges are indicated within `( )'. }
\label{tbl:rq2}
\small
\begin{tabular}{llr|r|r|r|r|r|r|r|r}
\rowcolor[HTML]{EFEFEF} 
\textbf{Policy} & \textbf{Classifier} & \textbf{Wins} & \textbf{Recall+} & \textbf{PF-} & \textbf{AUC+} & \textbf{D2H-} & \textbf{Brier-} & \textbf{G-Score+} & \textbf{IFA-} & \textbf{MCC+} \\ \hline
\early $E_{size}$ & LR &  & 75 (40) & \hil 28 (25) & \hil 70 (15) & \hil 34 (17) & \hil 30 (17) & 70 (25) & \hil 1 (3) & \hil 34 (26) \\
\early $E_{size}$ & SVM &  & 73 (44) & \hil 30 (27) & \hil 69 (15) & \hil 35 (19) & \hil 30 (18) & 69 (26) & \hil 1 (3) & \hil 33 (27) \\
- & $ManualDown$ & \multirow{-3}{*}{6} & \hil 86 (25) & {\color[HTML]{FE0000} \textbf{48 (15)}} & \hil 70 (12) & \hil 36 (11) & 41 (17) & \hil 75 (14) & \hil 1 (4) & \hil 30 (22) \\ \hline
\early $E_{size}$ & KNN & 4 & 70 (41) & \hil 33 (26) & \hil 67 (17) & \hil 37 (19) & 33 (18) & 66 (26) & \hil 1 (3) & 28 (28) \\ \hline
\early $E_{size}$ & NB &  & 75 (53) & \hil 38 (41) & 63 (17) & 44 (23) & 36 (25) & 61 (35) & \hil 1 (4) & 24 (31) \\
\early E & LR &  & 70 (42) & \hil 37 (33) & 63 (16) & 41 (22) & 36 (20) & 62 (30) & \hil 1 (4) & 24 (28) \\
\early E & SVM &  & 67 (42) & \hil 33 (31) & 64 (17) & 40 (20) & 34 (19) & 63 (30) & \hil 1 (4) & 25 (28) \\
\early E & KNN &  & 66 (43) & \hil 38 (30) & 62 (18) & 42 (20) & 37 (20) & 59 (28) & \hil 1 (4) & 21 (31) \\
\early $E_{size}$ & RF &  & 64 (40) & \hil 31 (30) & 64 (17) & 40 (20) & 33 (18) & 61 (28) & \hil 1 (4) & 25 (30) \\
\early $E_{size}$ & DT &  & 62 (39) & \hil 38 (30) & 61 (18) & 43 (19) & 38 (20) & 57 (28) & \hil 1 (4) & 19 (32) \\
\early E & RF &  & 60 (46) & \hil 32 (33) & 61 (18) & 43 (21) & 35 (21) & 56 (33) & \hil 1 (4) & 20 (32) \\
\early E & DT &  & 56 (46) & \hil 41 (37) & 56 (17) & 49 (22) & 41 (24) & 52 (33) & \hil 2 (5) & 11 (30) \\
\early E & NB & \multirow{-9}{*}{2} & 50 (75) & \hil 27 (48) & 54 (14) & 55 (30) & 36 (26) & 38 (62) & \hil 2 (7) & 8 (26) \\ \hline
- & $ManualUp$ & 0 & 14 (25) & 52 (15) & 30 (12) & 73 (11) & 59 (17) & 16 (27) & 7 (13) & -30 (22)
\end{tabular}%
\end{table*}

\begin{tcolorbox}[colback=red!5,colframe=gray!50]{\textit{At least for software defect prediction, early data with two features `la' and `lt' is \textit{better} than using recent (or more) data with many features.}}
\end{tcolorbox}

\subsection{\textbf{RQ3: Can we build early software defect prediction models from transferring early life cycle data from other projects?}} \label{tion:rq3}

\subsubsection{Motivation} \tion{sampling_methods} lists both the situations and challenges of when one might need to transfer project data to build defect predictors.  Although we cannot solve all \textit{Cross-project} challenges in one article, we could pacify some of them by checking the efficacy of early methods in this context. If early methods work, then we can transfer relevant data from vast search space (many projects) in less time. Further, we can also reduce the need for data availability as we will only need a small portion of early data. 

\subsubsection{Approach}

To check the efficacy of \textit{Cross-project} methods against early methods we will build predictors using the following sampling strategies below:
\bi
\item \textit{Cross-project} methods $Bellwether$ and $TCA+$ 
\item We created early variants of \textit{Cross-project} methods $Bellwether$ and $TCA+$ namely $E_{size} (Bellwether)$ and $E_{size} (TCA+)$. The details of these sampling strategies are elucidated in \tbl{policies} and portrayed visually in \fig{policies}.

\item We also include the within-project $E_{size}$ to compare \textit{Cross} and \textit{Within}. 

\item Note: We do not include $ALL$ or $E$ because $E_{size}$ outperformed those sampling strategies in  \tbl{rq2} and in our prior result~\cite{icse21}. 

\ei

The \textit{Cross-projects} methods require a representative project among the 240 projects to transfer data from. 

\begin{figure}[!t]
\begin{center}
  \includegraphics[width=2.25in,keepaspectratio]{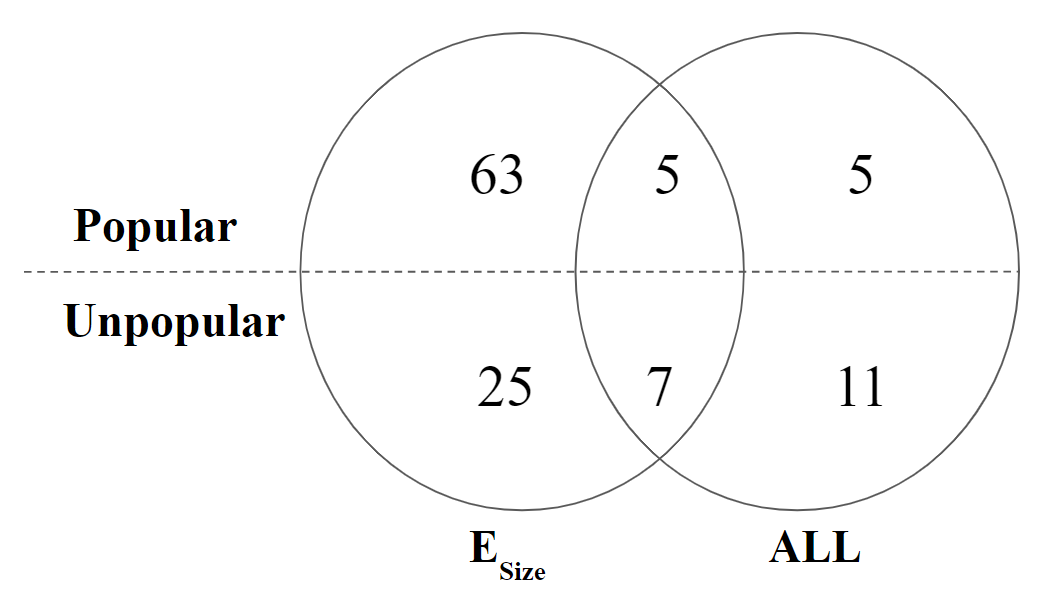}
\caption{Number of ``satisfactory''
bellwethers identified independently by two different sampling policies in all 240 projects.}\label{fig:commit_venn}  
\end{center}
\end{figure}

\subparagraph{Finding Representative Cross-Project:} 
Using just LA and LT for each of our projects, we found a model and measured 
its performance on the other  projects.
If that median performance was more the 70\% `Recall' and less than 30\% `PF', then the model was deemed ``satisfactory''.
In a result that endorses the use of early
life cycle data, Figure~\ref{fig:commit_venn}  
shows that {\em more} of the models found via
the $E_{size}$ method was ``satisfactory'' than those found using all the data. That effect is particularly marked in the popular projects~\footnote{Clearly, this analysis might be overly dependent on the ``magic numbers'' used to select ``satisfactory'' bellwethers, i.e., 70\% `Recall' and less than 30\% `PF.' But we have run our experiments using 12 ($i <= 12$) different bellwether projects with nearly equal frequency (see the seven unpopular + 5 popular bellwethers in the middle of Figure~\ref{fig:commit_venn}). The statistical analysis of the twelve $E_{size}$ (Bellwethers)  using \tion{sk} shows that (a)~they tie with each other and (b)~perform as well or \textit{better} than the other transfer learning methods explored in this paper.}.

Lastly like in previous RQs we compare the predictive performance of the various defect predictors that were built using different sampling policies  on all eight evaluation measures listed in \tion{measures} and rank them using the Scott-Knott test elucidated in \tion{sk}.

Note: We avoid the methodological error of testing the representative \textit{Cross-project} in this experiment.

\subsubsection{Results}

\begin{figure}[!t]
\begin{center}
\includegraphics[width=3in,keepaspectratio]{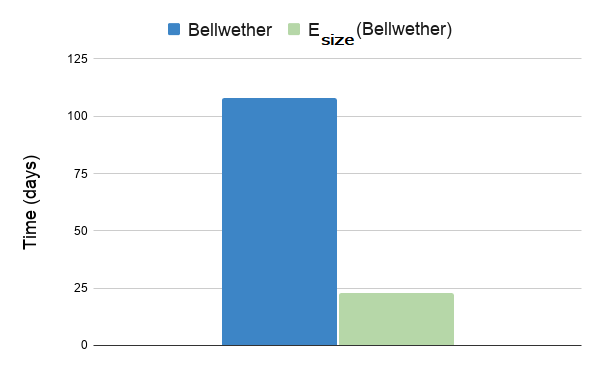}
\caption{Time taken (log-scale) to identify qualifying bellwether projects in all 240 projects using $ \mathit{Bellwether}$  and $ \mathit{E_{size} (Bellwether)}$ policies.
Note that in our experiments, we ran on a multi-core cloud-based CPU farm. To collect the data here (which assumes we are running on a single-core machine), we summed the CPU time across all the cores in our experimental rig.  We assert that this sum is valid since, in our experiments, there is nearly no communication between the cores (except that very end to accumulate the results).}\label{fig:img_runtime_rq3a}  
\end{center}
\end{figure}

\begin{figure*}[!b]
\begin{center}
\includegraphics[width=4in]{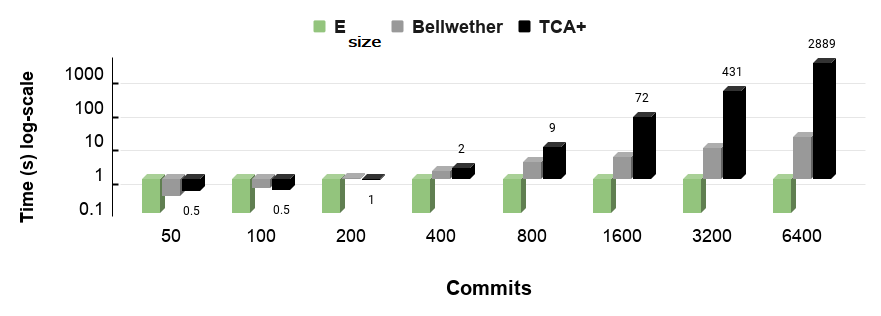}
\end{center}
\caption{Time taken by different sampling policies to train a model based on the number of commits.}\label{fig:img_runtime_rq3b} 
\end{figure*}

We make the following observations using \tbl{rq3}, \fig{img_runtime_rq3a} and \fig{img_runtime_rq3b}.

\bi

\item Rows \#1 and \#2 top of the \tbl{rq3} confirm that it is \textit{better} to build \textit{Cross} based predictors using early-data-lite method $E_{size}$ than using sampling strategies that use more data (like $TCA+$ or $Bellwether$). 

\item \fig{img_runtime_rq3a}  and \fig{img_runtime_rq3b} shows that it is much faster to identify and  build bellwethers from a large pool of projects using $E_{size}$ than $Bellwether$. This is because $E_{size}$ based predictors need not update (re-train) by accumulating newer project commits for every new project release.

\item Row \#6 of \tbl{rq3} is the position of our prior  result~\cite{icse21} that builds predictors using \textit{Within} $E$ project commits. That proves to show that we can do \textit{better} than  local data if we find good bellwethers.
\ei



\begin{table*}[h]
\caption{ `30' \textit{cross-project} and `1' \textit{within-project} (denoted as $\implies$) software defect prediction models tested in all 240 project's releases.
 In the first row, ``+'' and ``-'' denote the criteria that need to be maximized or minimized, respectively. 
 All the other  rows show combinations of sampling policies
 and classifiers.
Green cells denote ``early-data'' sampling was employed. Cells marked in 
gray all have the same top rank as the
best results (and those
ranks were determined by the Scott-Knott algorithm
described in \S\ref{tion:sk}).
The
`wins' column (see column \#3) counts how often a particular sampling policy/classifier achieves
a top score. Inter-quartile ranges are indicated within `( )'. }
\label{tbl:rq3}
\small
\begin{tabular}{llr|r|r|r|r|r|r|r|r}
\rowcolor[HTML]{EFEFEF} 
\textbf{Policy} & \textbf{Classifier} & \textbf{Wins} & \textbf{Recall+} & \textbf{PF-} & \textbf{AUC+} & \textbf{D2H-} & \textbf{Brier-} & \textbf{G-Score+} & \textbf{IFA-} & \textbf{MCC+} \\ \hline
\early $E_{size}$ (Bellwether) & SVM &  & \hil 91 (26) & 40 (34) & \hil 70 (16) & \hil 35 (21) & 35 (23) & \hil 74 (23) & \hil 1 (4) & \hil 32 (26) \\
\early $E_{size}$ (TCA+) & SVM &  & \hil 88 (33) & 38 (28) & \hil 70 (16) & \hil 35 (19) & 34 (21) & \hil 74 (24) & \hil 1 (4) & \hil 32 (26) \\
\early $E_{size}$ (Bellwether) & LR & \multirow{-3}{*}{6} & \hil 85 (30) & 34 (27) & \hil 73 (15) & \hil 31 (17) & 31 (18) & \hil 76 (20) & \hil 1 (4) & \hil 35 (25) \\ \hline
\early $E_{size}$ (TCA+) & LR &  & 82 (36) & 32 (25) & \hil 72 (16) & \hil 33 (19) & 31 (19) & \hil 74 (22) & \hil 1 (4) & \hil 34 (26) \\
\early $E_{size}$ (TCA+) & KNN &  & 79 (50) & 32 (25) & \hil 70 (18) & \hil 35 (20) & 32 (18) & \hil 72 (28) & \hil 1 (5) & \hil 30 (28) \\
\early $\implies$ $E_{size}$ (Within) & LR &  & 78 (43) & 25 (25) & \hil 73 (16) & \hil 31 (18) & 27 (18) & \hil 73 (28) & \hil 1 (3) & \hil 37 (27) \\
\early $E_{size}$ (Bellwether) & KNN &  & 78 (43) & 30 (26) & \hil 70 (17) & \hil 34 (19) & 30 (19) & \hil 72 (27) & \hil 1 (3) & \hil 32 (27) \\
\early $E_{size}$ (TCA+) & RF & \multirow{-5}{*}{5} & 75 (46) & 31 (24) & \hil 69 (17) & \hil 35 (19) & 32 (18) & \hil 71 (28) & \hil 1 (4) & \hil 30 (28) \\ \hline
\early $E$ (TCA+) & LR & 4 & 80 (40) & 41 (19) & \hil 69 (16) & \hil 36 (15) & 38 (17) & \hil 72 (24) & \hil 2 (5) & 28 (27) \\
\early $E_{size}$ (Bellwether) & RF &  & 71 (41) & 28 (26) & 68 (18) & \hil 36 (21) & 30 (19) & 67 (28) & \hil 1 (4) & \hil 28 (28) \\
\bellwether Bellwether & SVM &  & 21 (44) & \hil 4 (10) & 57 (17) & 56 (30) & \hil 19 (15) & 25 (49) & \hil 1 (7) & 23 (40) \\
\bellwether Bellwether & RF & \multirow{-3}{*}{3} & 19 (36) & \hil 3 (8) & 56 (15) & 56 (25) & \hil 19 (15) & 23 (41) & \hil 1 (7) & 22 (39) \\ \hline
\early $E_{size}$ (Bellwether) & NB &  & \hil 88 (33) & 49 (34) & 64 (19) & 42 (22) & 42 (23) & 67 (28) & \hil 2 (5) & 23 (27) \\
\early $E_{size}$ (TCA+) & NB &  & 82 (38) & 43 (30) & 67 (18) & 39 (21) & 38 (22) & \hil 70 (26) & \hil 2 (5) & 25 (28) \\
\bellwether Bellwether & LR &  & 76 (51) & 33 (43) & 67 (16) & 39 (20) & 32 (24) & 66 (34) & \hil 1 (4) & \hil 31 (28) \\
\early $E_{size}$ (TCA+) & DT &  & 71 (39) & 32 (25) & 67 (19) & \hil 37 (19) & 33 (18) & 67 (27) & \hil 2 (4) & 27 (28) \\
\bellwether Bellwether & KNN & \multirow{-5}{*}{2} & 25 (41) & 6 (10) & 59 (17) & 53 (27) & \hil 20 (16) & 28 (44) & \hil 1 (5) & 24 (39) \\ \hline
\early $E$ (TCA+) & KNN &  & 71 (53) & 37 (28) & 64 (20) & 40 (21) & 36 (20) & 65 (33) & \hil 2 (5) & 23 (28) \\
\early $E$ (TCA+) & SVM &  & 70 (55) & 34 (31) & 64 (20) & 41 (24) & 35 (21) & 64 (40) & \hil 2 (5) & 23 (32) \\
\tca TCA+ & LR &  & 70 (39) & 50 (22) & 61 (22) & 42 (21) & 45 (22) & 64 (28) & \hil 2 (6) & 17 (35) \\
\early $E$ (TCA+) & NB &  & 67 (56) & 39 (38) & 60 (20) & 46 (24) & 38 (24) & 60 (44) & \hil 2 (6) & 17 (32) \\
\early $E_{size}$ (Bellwether) & DT &  & 67 (44) & 30 (26) & 65 (20) & 39 (23) & 33 (19) & 63 (32) & \hil 1 (4) & 25 (32) \\
\early $E$ (TCA+) & DT &  & 67 (42) & 37 (25) & 63 (19) & 41 (20) & 38 (18) & 62 (30) & \hil 2 (5) & 20 (30) \\
\early $E$ (TCA+) & RF &  & 67 (40) & 34 (23) & 65 (20) & 38 (19) & 35 (18) & 65 (28) & \hil 2 (5) & 24 (28) \\
\bellwether Bellwether & NB &  & 50 (60) & 15 (21) & 64 (22) & 41 (32) & 25 (16) & 53 (55) & \hil 1 (5) & 26 (38) \\
\tca TCA+ & SVM &  & 50 (50) & 23 (27) & 61 (21) & 45 (25) & 31 (21) & 52 (41) & \hil 2 (5) & 20 (34) \\
\tca TCA+ & RF &  & 50 (50) & 31 (32) & 57 (17) & 48 (23) & 36 (23) & 49 (36) & \hil 2 (6) & 13 (28) \\
\tca TCA+ & DT &  & 50 (40) & 34 (28) & 56 (17) & 49 (21) & 39 (21) & 48 (31) & \hil 2 (6) & 11 (28) \\
\tca TCA+ & KNN &  & 40 (42) & 19 (22) & 57 (18) & 49 (24) & 28 (21) & 42 (36) & \hil 2 (5) & 16 (32) \\
\bellwether Bellwether & DT & \multirow{-13}{*}{1} & 33 (33) & 13 (12) & 59 (16) & 48 (21) & 24 (16) & 38 (34) & \hil 1 (5) & 19 (32) \\ \hline
\tca TCA+ & NB & 0 & 23 (47) & 19 (35) & 50 (14) & 61 (22) & 34 (28) & 23 (48) & 3 (9) & 0 (28)
\end{tabular}%
\end{table*}

\begin{tcolorbox}[colback=red!5,colframe=gray!50] \textit{Transfer Learning methods perform better (faster and more accurate) when they are sampled using the early method. }
\end{tcolorbox}

\subsection{\textbf{RQ4: Do complex methods supersede early software defect prediction models ?}}

\subsubsection{Motivation} Our prior work and RQs 1, 2, and 3 have used classifiers with default parameter settings (off the shelf). However, numerous studies, especially in the space of software defect prediction, have shown considerable improvement in predictive performance when the classifiers are tuned~\cite{fu2016tuning,agrawal2019dodge} or using an ensemble approach~\cite{yang2017tlel}. But note the downside of tuning is the run-time overhead. 

Therefore, it is essential to check if complex methods like tuning or ensemble favor sampling policies that use recent (or more) data than those confined to the early regions. It is also motivating to check if early methods can benefit from these complex methods. Perhaps the run-time overheads can be alleviated by working with fewer early data.

\subsubsection{Approach}
We build predictors using the state-of-the-art methods, specifically $DODGE$, $Hyperopt$, and $TLEL$ elucidated in \tion{learners}.  RQ's 1 to 3  in either  \textit{Within} or \textit{Cross} context $E_{size}$ has performed \textit{similar} or \textit{better} than prevalent sampling strategies, indicating the importance of early regions of the software project. Therefore like RQ\#3 we create variants of $DODGE$ , $Hyperopt$ and $TLEL$ using $E_{size}$. In other words, we will input fewer early data $E_{size}$ and use those complex methods to check if they perform better or worse. 

The optimizers $DODGE$ and $Hyperopt$ find the best set of hyper-parameters using the commits sampled according to the policy rule (specified in \tbl{policies})  to minimize $D2H$ score (elucidated in \tion{measures}). We do not sample tuning commits separately; the optimizers sample the tuning commits within the training commit boundary defined by the sampling policy. We note here that we tried to use 50\% of the training commits for training and the rest for tuning and used 100\% of the commits for both training and tuning; neither stratification impacted the optimizer results.

Since the optimizers have a huge run-time overhead, and our sample includes projects with over 10,000 commits (for e.g., homebrew-cask has 98,362 commits). Therefore, it is not feasible to test many large models repeatedly on over 10,000+ releases across multiple classifiers. On the other hand, data should not be scarce for optimizers to learn tuning parameters effectively. Hence, we limit optimizers to use up to 957 commits (in the case of $ALL$). This is because, in, RQ1 we confirmed that the project's popularity has no impact on the results. Hence, we use the median commits (957) of the unpopular projects (see \fig{distribution_unpopular}), which we think is a reasonable representation of the whole population. 

Lastly, we measure the predictive performance of the above approaches listed in \tion{measures} and  rank the predictive performance of defect predictors that were built using the Scott-Knott test elucidated in \tion{sk}.

\subsubsection{Results}

We make the following observations from \tbl{rq4}:

\bi
\item $E_{size}$ (row \#1) performs \textit{better} than all complex methods that use either more data $ALL$ or fewer early date $E_{size}$.

\item Closest to $E_{size}$ is the two-layer ensemble method that achieves almost \textit{similar} results but with more data and more processing. But note $TLEL$ (row \#2) was trained on every new project release accumulating more data, whereas $E_{size}$ or any $E$ based method is always trained just `once'.
\ei

\begin{table*}[h]
\caption{ 7 \textit{within-project} software defect prediction models tested in all 240 project releases.
 In the first row, ``+'' and ``-'' denote the criteria that need to be maximized or minimized, respectively. 
 All the other  rows show combinations of sampling policies
 and classifiers.
Green cells denote ``early-data'' methods (and all other rows employ an optimizer or ensemble approach). Cells marked in
gray all have the same top rank as the
best results (and those
ranks were determined by the Scott-Knott algorithm
described in \S\ref{tion:sk}).
The
`wins' column (see column \#3) counts how often a particular sampling policy/classifier achieves
a top score. Inter-quartile ranges are indicated withing `( )'.}
\label{tbl:rq4}
\small
\begin{tabular}{llr|r|r|r|r|r|r|r|r}
\rowcolor[HTML]{EFEFEF} 
\textbf{Policy} &
  \textbf{Classifier} &
  \textbf{Wins} &
  \textbf{Recall+} &
  \textbf{PF-} &
  \textbf{AUC+} &
  \textbf{D2H-} &
  \textbf{Brier-} &
  \textbf{G-Score+} &
  \textbf{IFA-} &
  \textbf{MCC+} \\ \hline
\early $E_{size}$ & LR                 & 7                   & \hil 73 (43)  & 28 (26) & \hil 70 (15) & \hil 34 (18) & \hil 30 (17) & \hil 69 (26) & \hil 1 (3) & \hil 35 (27) \\ \hline
ALL        & TLEL     &                     & \hil 74 (50)  & 36 (56) & 61 (20)      & 46 (31)      & 36 (31)      & 54 (46)      & \hil 1 (4) & 21 (38)      \\
\early $E_{size}$ & TLEL     &                     & \hil 67 (38)  & 33 (30) & 64 (17)      & 39 (19)      & 34 (19)      & 62 (27)      & \hil 1 (4) & 26 (30)      \\
\early $E_{size}$ &
  HyperOpt &
  \multirow{-3}{*}{2} &
  27 (60) &
  \hil 11 (31) &
  54 (14) &
  56 (28) &
  \hil 28 (22) &
  28 (55) &
  2 (11) &
  10 (28) \\ \hline
\early $E_{size}$ & DODGE    &                     & \hil 100 (80) & 68 (91) & 50 (7)       & 71 (17)      & 49 (46)      & 0 (42)       & 3 (9)      & 0 (18)       \\
ALL        & DODGE    & \multirow{-2}{*}{1} & \hil 100 (80) & 98 (69) & 50 (0)       & 71 (3)       & 66 (36)      & 0 (17)       & 4 (9)      & 0 (0)        \\ \hline
ALL        & HyperOpt & 0                   & 50 (89)       & 40 (66) & 50 (9)       & 63 (21)      & 44 (30)      & 27 (53)      & 3 (9)      & 0 (21)      
\end{tabular}%
\end{table*}

\begin{figure*}[h]
\begin{center}
\includegraphics[width=3.25in]{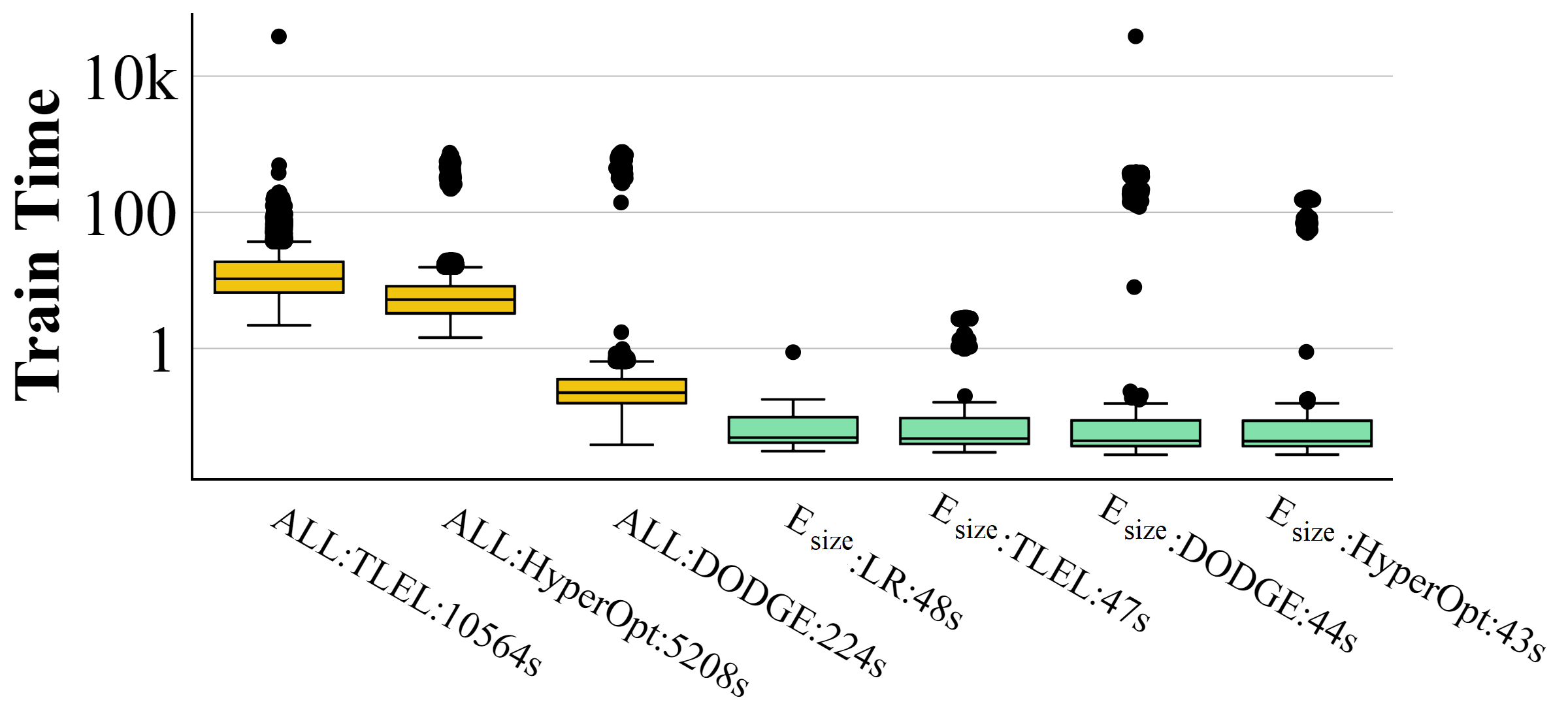}
\end{center}
\caption{Time taken by different sampling policies to train a model based on the number of commits.}\label{fig:img_runtime_rq4} 
\end{figure*}

Notably, \fig{img_runtime_rq4} confirms it is far more faster to build predictors sampled using $E_{size}$ that neither needs SMOTE or CFS and only trained once throughout the project life cycle.

\begin{tcolorbox}[colback=red!5,colframe=gray!50]{\textit{Neither Optimizers nor ensemble methods surpass simpler early methods.}}
\end{tcolorbox}

\section{Threats to Validity}\label{tion:threats}

\subsection{Sampling Bias} Generalizability of the conclusions will rely on the examples considered; i.e., what is essential here may not be genuine all over. Although the prevalent practice of such empirical studies is to use popular OS GitHub projects, we broaden the scope by including unpopular projects. Nevertheless, all these projects are non-trivial engineering projects developed in numerous programming languages for various domains. Notably, lessons did not vary in either of these two populations of projects. 

\subsection{Construct Validity}

We mined the project samples using \textit{Commit-Guru}. \textit{Commit-Guru} determines risky Commits using a method similar to  SZZ~\cite{cite_szz_original}. Such an approach despite its prevalence is widely debated for its false positives. To minimize that threat \textit{Commit-guru} only considers files that are source-code files and not documents like readme or pdf files. \textit{Commit-Guru} classifies commit into different categories one of them is \textit{merge-commit}. Studies have shown merge commits have little to no influence over making real changes to the software, therefore in this study, we filter all merge commits. 

\begin{figure*}[!t]
\begin{center}
\includegraphics[width=\linewidth]{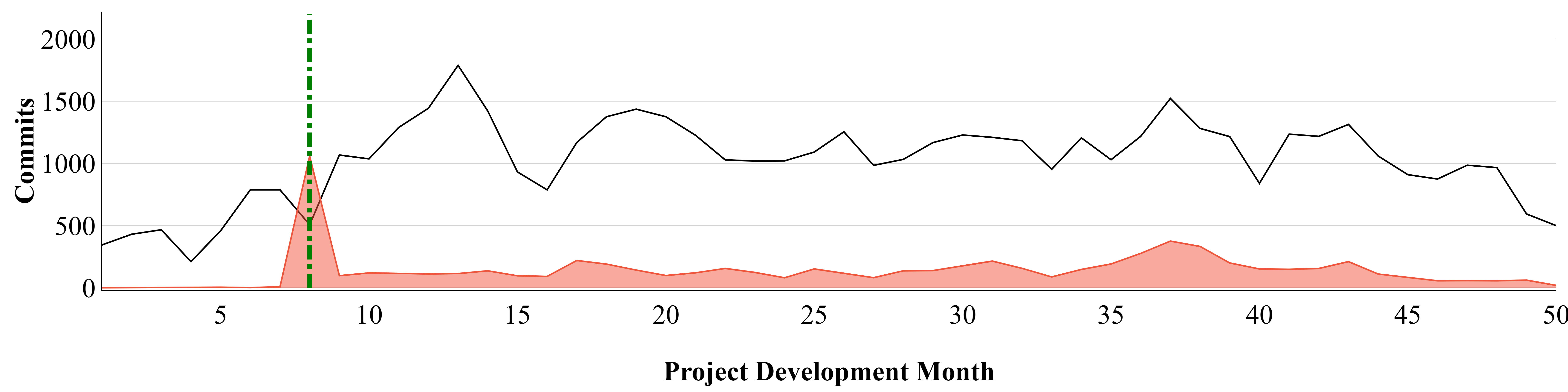}
\end{center}
\caption{A repeated pattern where  most defects (shaded in red) occur early in the life cycle (before the green dotted line) is also observed among the two  projects (QT and OPENSTACK systems) with  37,524 commits  assessed recently by McIntosh and Kamei in ~\cite{mcintosh2017fix}.}\label{fig:whale_threat}
\end{figure*}

Lastly, to further bold the validity of our early effect we cross-checked our results with the two projects `QT' and `OPENSTACK' used in the 2017 TSE article by McIntosh and Kamei~\cite{mcintosh2017fix} and found our early trend and effect to hold. Please see the early trend of those two projects in \fig{whale_threat} and the results in \tbl{threat}. \tbl{threat} shows results of defect predictors that sampled training commits using the recent six months (M6) endorsed in ~\cite{mcintosh2017fix} versus $E_{size}$ endorsed in this study. Note our results are compared with ten classifiers and eight evaluation measures.

\begin{table*}[h]
\caption{ `12' \textit{within-project}  software defect prediction models tested `only' in two projects `QT' and `OPENSTACK'.
 In the first row, ``+'' and ``-'' denote the criteria that need to be maximized or minimized, respectively. 
 All the other  rows show combinations of sampling policies
 and classifiers.
Green cells denote ``early-data'' sampling was employed. Cells marked in 
gray all have the same top rank as the
best results (and those
ranks were determined by the Scott-Knott algorithm
described in \S\ref{tion:sk}).
The
`wins' column (see column \#3) counts how often a particular sampling policy/classifier achieves
a top score. Inter-quartile ranges are indicated within `( )'. }
\label{tbl:threat}
\small
\begin{tabular}{llr|r|r|r|r|r|r|r|r}
\rowcolor[HTML]{EFEFEF} 
\textbf{Policy} & \textbf{Classifier} & \textbf{Wins} & \textbf{Recall+} & \textbf{PF-} & \textbf{AUC+} & \textbf{D2H-} & \textbf{Brier-} & \textbf{G-Score+} & \textbf{IFA-} & \textbf{MCC+} \\ \hline
\early $E_{size}$ & RF & 6 & \hil 71 (13) & 35 (8) & \hil 67 (7) & \hil 33 (6) & 35 (8) & \hil 69 (10) & \hil 5 (7) & \hil 19 (9) \\ \hline
\early $E_{size}$ & NB & 5 & \hil 75 (14) & 43 (7) & \hil 67 (9) & 35 (7) & 42 (8) & \hil 71 (9) & \hil 6 (8) & \hil 17 (10) \\ \hline
\early $E_{size}$ & SVM &  & 67 (18) & 33 (9) & \hil 67 (7) & \hil 34 (8) & 33 (8) & 66 (14) & \hil 4 (7) & \hil 18 (11) \\
\early $E_{size}$ & LR & \multirow{-2}{*}{4} & 64 (16) & 27 (7) & \hil 67 (8) & \hil 33 (8) & 28 (6) & 65 (12) & \hil 4 (8) & \hil 21 (11) \\ \hline
$M6$ & SVM &  & 14 (25) & \hil 6 (5) & 54 (10) & 61 (17) & \hil 13 (7) & 17 (28) & \hil 3 (8) & 9 (16) \\
$M6$ & RF & \multirow{-2}{*}{3} & 13 (19) & \hil 4 (5) & 55 (8) & 61 (13) & \hil 9 (10) & 16 (22) & \hil 3 (8) & 11 (17) \\ \hline
$M6$ & KNN & 2 & 20 (13) & 8 (6) & 56 (7) & 56 (9) & \hil 13 (8) & 24 (14) & \hil 3 (7) & 11 (13) \\ \hline
\early $E_{size}$ & DT &  & 67 (18) & 48 (6) & 59 (7) & 42 (6) & 47 (5) & 63 (11) & \hil 7 (9) & 8 (10) \\
\early $E_{size}$ & KNN &  & 65 (17) & 35 (18) & 65 (7) & 36 (7) & 35 (16) & 65 (13) & \hil 5 (9) & 16 (10) \\
$M6$ & DT &  & 25 (21) & 12 (12) & 55 (7) & 54 (12) & 15 (14) & 28 (22) & \hil 4 (11) & 8 (11) \\
$M6$ & NB & \multirow{-4}{*}{1} & 14 (22) & 9 (12) & 52 (8) & 62 (12) & 17 (11) & 17 (21) & \hil 6 (13) & 5 (13) \\ \hline
$M6$ & LR & 0 & 10 (28) & 10 (13) & 51 (6) & 64 (17) & 17 (12) & 12 (28) & 9 (52) & 1 (10)
\end{tabular}%
\end{table*}

\subsection{Learner bias} For identifying bellwethers by all-pairs experiment (each project is tested on all 12,000+ releases), we used only one classifier, `Logistic Regression.' Perhaps other classifiers may qualify other projects are bellwethers. Using just one classifier may not be a threat for the following two reasons. Firstly a Logistic-Regression is recommended in the baseline study~\cite{icse21} and widely endorsed in classifiers in software defect prediction literature. Secondly, in all our results in \tion{results} Logistic-Regression based predictors were in-par and \textit{better} than five other classifiers explored in this study.

Further, an empirical study can only focus on a handful of representative classifiers.  Thus we chose ten classifiers (Logistic Regression, Nearest neighbor, Decision Tree, Random Forrest, and Na\"ive Bayes). These ten classifiers cover a broad range of classification algorithms~\cite{ghotra2015revisiting}.

\subsection{Evaluation bias} 
This paper uses both `Recall' and `PF' to identify bellwethers and eight evaluation measures (Recall, PF, IFA, Brier, GM, D2H, MCC, and AUC) to compare the policies extensively. Other widely used measures in software defect prediction are precision and f-measure. However, as mentioned earlier, those threshold-dependant metrics have issues with unbalanced data~\cite{menzies2008implications}.

\subsection{Input Bias} All our proposed sampling policies randomly sample 50 commits from the first 150 commits of the project. Along these lines, it could be true that different executions could yield different results. However, this is not a threat, as our conclusions hold on a large sample size of 12,000+ releases.

\section{Discussion}\label{tion:discussion}

\begin{figure}[!t]
\begin{center}
   \includegraphics[width=4.5in,keepaspectratio]{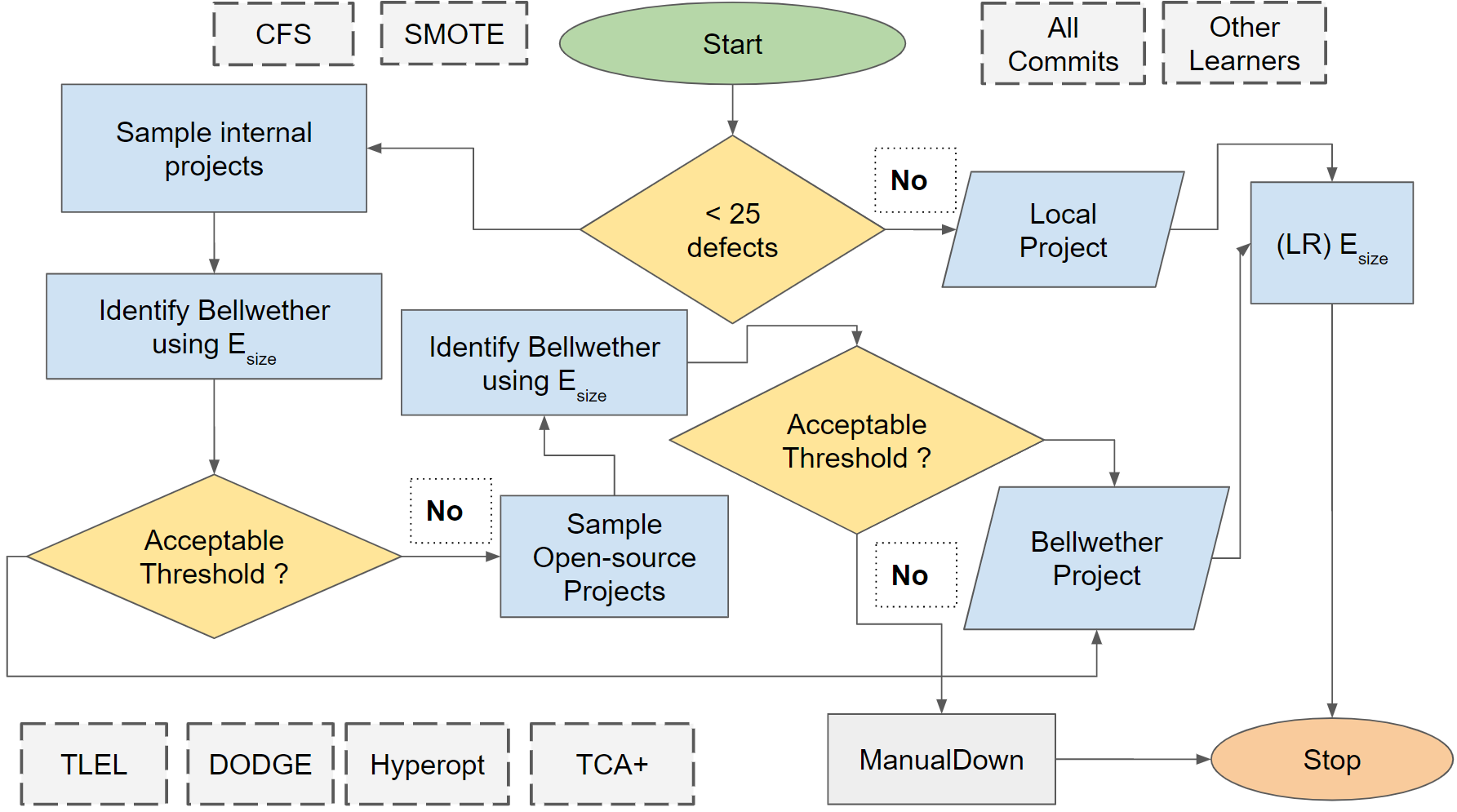}
\caption{A decision tree to build defect predictors drawn using the  results from \tion{results} that find  many complex methods (shown as disconnected blocks with dashed borders) to be needless.}\label{fig:img_flowchart}  
\end{center}
\end{figure}

We summarized the results from RQ1 to RQ4 as a simple decision tree drawn in \fig{img_flowchart}. The first rule in the tree eliminated the data-hungry approaches listed in \tbl{lit_review} by showing that the project under test only requires 25 clean and 25 defective commits to predict post-release defects. The results of this work endorse that practitioners can  quickly build an un-tuned $E_{size}$ and $LR$ defect predictor based on two size-based features. More importantly, researchers should use the early-bird treatment as a baseline before trying sophisticated methods.

In the absence of defective commits, the decision tree suggests seeking other projects from the same organization to find a suitable $Bellwether$ project. While $Bellwether$ is not a novel technique, the brute-force process of finding bellwethers can be very expensive as shown in \fig{img_runtime_rq3a}. This work pacifies the O($n^2$) process by only looking at the first 150 commits (constant) in each of the bellwether project populations of size `n'. If other projects from the same organization are unavailable due to issues such as data sensitivity, the tree endorses the use of open-source projects  to find a suitable bellwether(s). 

In situations when there is no access to data or the inability to find good bellwether projects, we suggest practitioners use $ManualDown$ until their local or other projects from the same organization report more defects ( $> 24$ ). Based on the results of the RQs we find the complex tuning/ensemble methods and data-hungry sampling methods were needless and thus they are disconnected from the main flow.

\subsection{Management Implications}\label{manage}
Much prior research has struggled to find stable conclusions that hold across multiple projects.
Menzies et al. warn that many global lessons (generalizations) are not supported locally~\cite{menzies2011local}. 
That is to say, the more data we see, the more exceptions
and special cases might appear in the models learned
from that data.
Such conclusion instability in SE has been  often recorded  in~\cite{zimmermann2009cross,menzies2011local,shrikanth2020assessing}.

Is that the best we can do? 
Ideally, SE research can offer stable general software defect prediction principles (such as those seen above)
to guide project management, software standards, education,   tool development, and legislation about software. 
 Such conclusion stability
would have benefits for   {\em   trust, insight, training}, and {\em tool
          development}.

{\em Trust:}  
Conclusion instability is unsettling for project managers. Hassan~\cite{hassan17} warns that managers lose trust in software analytics if its results keep changing.
But if we can find stable conclusions (e.g., using early life cycle bellwethers),
that would give   project managers   clear guidelines on many issues, including (a) when a certain module should be inspected; (b) when modules should be refactored; 
and (c) deciding where to focus on expensive testing procedures.
 
{\em Insight:} 
Sawyer et al. assert that insights are essential to catalyzing business initiative~\cite{sawyer2013bi}. 
From Kim et al.~\cite{Kim2016} perspective, software analytics is a way to obtain fruitful insights that guide practitioners to accomplish software development goals, whereas for Tan et al.~\cite{tan2016defining} such insights are a central goal. From a practitioner's perspective, Bird et al.~\cite{Bird:2015} report, insights occur when users respond to software analytics models. Frequent model regeneration can exhaust users' ability to confident conclusions from new data. In this regard, we note that the early-bird models
found would be a stable source of insight for much of the project life cycle.
 
{\em Tool development and Training:} 
Previously~\cite{shrikanth2020assessing} we warned that unstable models make it hard to onboard novice software engineers. Without knowing what factors most influence the local project, it is hard to
design and build appropriate tools for quality assurance activities. Hence, here again, the stability of our early life cycle detectors is very useful.

\begin{figure}[!t]
\begin{center}
\includegraphics[width=2.6in]{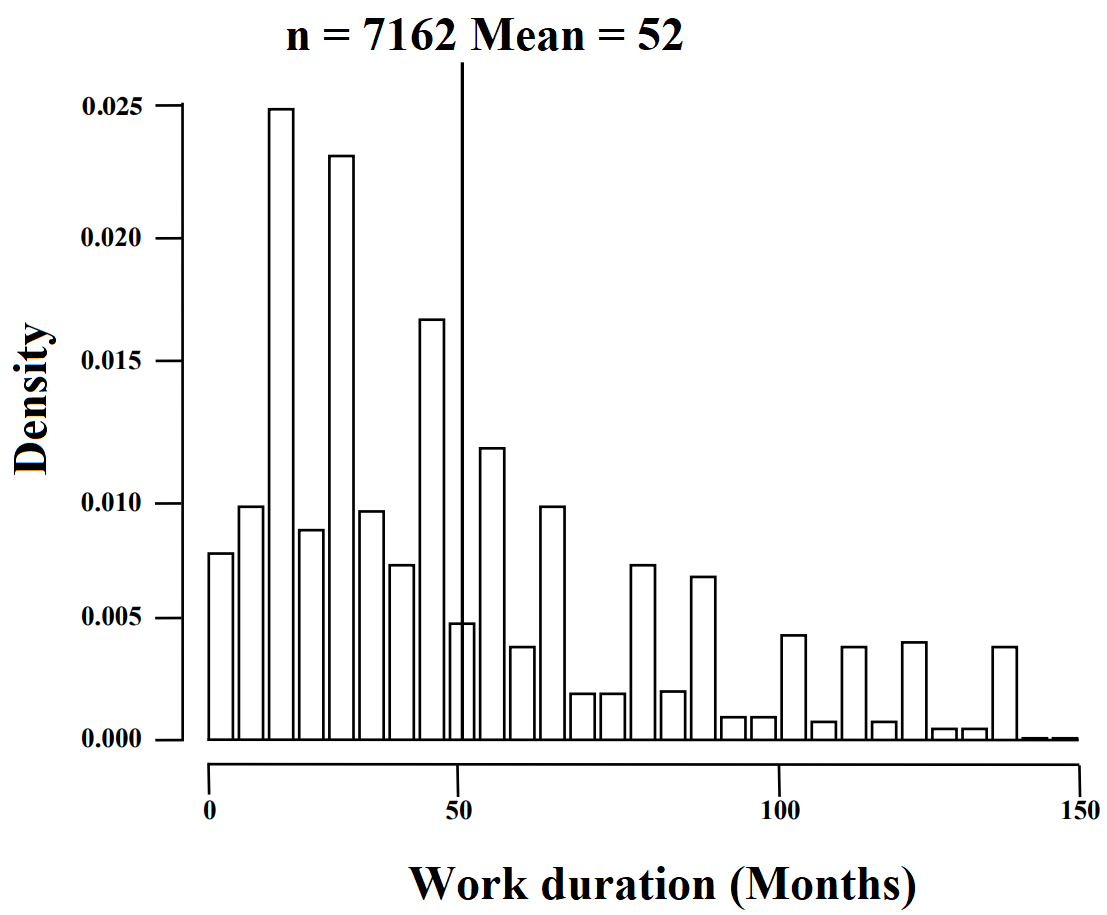}
\end{center}
\caption{Work duration histograms on particular projects;  from~\cite{Sela18}. Data from:  Facebook, eBay, Apple, 3M, Intel, and Motorola. }\label{fig:work}  
\end{figure}

All these problems with  trust, insight, training, and tool
          development
          can be solved if, early on in the project, a  software defect prediction model can be learned that is effective for the rest of the life cycle (which is the main result of this paper).
 Within that data, 
 we have found that  
 models learned
after just     150 commits, perform just as well
as anything else. In terms of resolving conclusion instability, this is a very significant result since
it means that for much of the life cycle, we can offer stable defect predictors.

One way to consider the impact of such early life cycle predictors is to use the data of \fig{work}.
That plot shows that software employees usually change projects every 52 months
(either moving between companies or changing projects within an organization).
According to \fig{work}, in seven years (84 months), the majority of workers and managers would first appear on a job 
{\em after}
the initial four months are required to learn a defect predictor.  Hence,  for most workers and managers,
the detectors learned via the methods of this paper would be the ``established wisdom'' and ``the way we do things here'' for their projects.
This means that a detector learned in the first four months would
be a suitable oracle to guide training and hiring;  the development of code review practices; the automation
of local ``bad smell detectors''; as well as tool selection and development.

\subsection{Systems Implications} 
\fig{img_runtime_rq3a}  shows the run-time required
to find  bellwether projects in all 240 projects using $ \mathit{Bellwether}$  or $ \mathit{E_{size} (Bellwether)}$ policies.
Note that the early-bird methods use 2.5 months less CPU than the alternative.

Since we only need the first 150 commits to identify a suitable project, many projects in their early stages may be included in the pool of transferable projects. 
Importantly \fig{commit_venn} show when sampled using $\mathit{E_{size}}$ 30\% ``more projects'' were qualified in much ``less time''. The choice of bellwether does not matter; perhaps, in practice, practitioners need not continue to look for bellwethers and terminate early as soon as they find the first bellwether. This could also mean lesser privacy concerns and enable practitioners to share fewer data freely. The early-data nature of our methods can also help to gauge sophisticated techniques like TCA+ faster. 

Note: We do not claim $E_{size} (Bellwether)$ is data-lite because it would still process all the projects in the search space. But we do endorse $E_{size}$ over current methods $Bellwether$ or $TCA+$. Because $E_{size}$   uses fixed data (150 commits) and does not re-train for every new project release it is faster and more accurate (see \tion{rq3}).

Practitioners may feel that retraining defect prediction models for every project release may not be a tedious activity. However, our prior work~\cite{icse21} points out two concerns. First, models that are frequently retrained may yield different conclusions. \respto{2N} \BLUE As seen from \fig{whale} information is rich in the earlier periods of the project. When including training data from uninformative regions (training with recent releases or all available data) the model would yield different conclusions. In other words, a  commit that was classified as defective could be classified as clean in another project release causing confusion within teams. \BLACK Second, it is a methodological error to learn (sample training data) from the shallow end of the project region.

\subsection{Research Implications}
We have shown that when defect data contains information, that information may be densest in a small part of the historical record of a project. 
While we have {\em not} shown that other kinds of SE data have the same density effect, we would argue that now
it is at least an open question that ``have we been learning from the wrong parts of the data?''.


Finally, we should now view it as a   potential
methodological error to reason across
all data in a project.
In the specific case of software defect prediction
we must now revisit any conclusion based just on later life cycle data
There are many examples of such conclusions. For example:
\begin{itemize}
\item
Hoang et al. says 
 ``We assume that older commits changes may have characteristics that no longer effects to the latest commits''~\cite{8816772}.
 \item
Also,
it is common practice in software defect prediction to perform ``recent  validation'' where predictors are tested on  the latest release after training from the prior one or two
 releases~\cite{tan2015online,mcintosh2017fix,kondo2020impact,fu2016tuning}. 
 \end{itemize}
More generally,  before researchers focus on later life cycle data, they must first check that their buggy commit data occur at equal frequency across the life cycle. If the buggy commits data is isolated to a specific region, then predictors should be built from that region. This work finds that most buggy commits are found early in the project life cycle.

\section{Conclusion} \label{tion:conclusion}

Issues with conclusion instability, run-time complexity, and explanations of models improve if we can learn a   predictive model early in the life cycle that is effective for the rest of the project. The prior work on software defect prediction identified the ``early bird'' heuristic. However, the results were scoped only in the context of within-project defect prediction on \textit{popular} GitHub projects~\cite{icse21}. 

We expanded the scope of the early method to simplify software defect prediction. Specifically, the results of this work find that a small early life cycle sample with two features can replace some methods and usefully augment others. For within-project defect prediction,  when trying to improve predictive performance,  the extra CPU required for (a) ensemble and (b) hyper-parameter optimization is not needed since sampling from early life cycle data performs \textit{better} than both (a) and (b).  As to transfer learning in cross-company learning, this work recommends that a small early life cycle sample on cross-project learning is significantly \textit{better} than using the entire cross-project life cycle data.

As for future work, perhaps we can simplify many SE tasks using the ``early bird'' heuristic.

\section*{Acknowledgements}

This work was partially supported by an NSF-CISE grant \#1908762.

\bibliographystyle{ACM-Reference-Format}
\bibliography{bib_references}

\end{document}